\documentclass[10pt]{article}
\usepackage{ametsoc2col}
\newcommand{\cD}{c_{_\mathrm{D}}}

\newcommand{\myabstract}{This paper discusses some basic
concepts that arise in the study of the tropical cyclone frictional boundary
layer. Part I discusses the concepts of asymptotic triangular waves
and asymptotic N-waves in the context of the nonlinear advection equation
and Burgers' equation. Connections are made between triangular waves
and single eyewalls, and between N-waves and double eyewalls. In Part II,
analytical solutions of a line-symmetric, $f$-plane, slab model of the
atmospheric boundary layer are presented. The boundary layer flow is
forced by a specified pressure field and initialized with $u$ and $v$
fields that differ from the steady-state Ekman solution. With certain
smooth initial conditions, discontinuities in $u$ and $v$ can be produced
during the transient adjustment to the steady-state Ekman solution.
Associated with these discontinuities in the horizontal wind components
are singularities in the boundary layer pumping and the boundary layer
vorticity, which can be either divergence-preferred or vorticity-preferred.
These models serve as a prototype for understanding the role of the
atmospheric boundary layer in the dynamics of primary and secondary
eyewalls in tropical cyclones.}

\begin{document}
%
\title{\textbf{\large{Basic Concepts Involved in Tropical Cyclone Boundary Layer Shocks}}}
\author{\textsc{Wayne H.\ Schubert,} \\
\and
\centerline{\textsc{Christopher J.\ Slocum, and Richard K. Taft}}\\
\centerline{\textit{\footnotesize{Department of Atmospheric Science,
Colorado State University, Fort Collins, Colorado, USA}}}
}
\ifthenelse{\boolean{dc}}
{
\twocolumn[
\begin{@twocolumnfalse}
\amstitle

\begin{center}
\begin{minipage}{13.0cm}
\begin{abstract}
	\myabstract
	\newline
	\begin{center}
		\rule{38mm}{0.2mm}
	\end{center}
\end{abstract}
\end{minipage}
\end{center}
\end{@twocolumnfalse}
]
}
{
\amstitle
\begin{abstract}
\myabstract
\end{abstract}
\newpage
}

\centerline{\bf Contents}

\addtolength{\leftmargini}{0.2in}
\setlength{\leftmarginii}{0.25in}
\begin{enumerate}
\setlength{\itemsep}{0pt}
\setlength{\parskip}{0pt}
\setlength{\parsep}{0pt}
\item[I.] Advection Equation and Burgers' Equation
\begin{enumerate}
\setlength{\itemsep}{0pt}
\setlength{\parskip}{0pt}
\setlength{\parsep}{0pt}
\item[1.] Introduction
\item[2.] Asymptotic triangular waves and their conceptual connection
      with primary eyewalls
\item[3.] Asymptotic N-waves and their conceptual connection with moats and double eyewalls
  \begin{enumerate}
     \item[3a.] Undamped N-waves
     \item[3b.] Damped N-waves
  \end{enumerate}
\item[4.] Triangular waves and primary eyewalls from Burgers' equation
\item[5.] N-waves, moats, and double eyewalls from Burgers' equation
\item[6.] Axisymmetric shocks
\end{enumerate}
\item[II.] Line-Symmetric Slab Ekman Layer Model
\begin{enumerate}
\setlength{\itemsep}{0pt}
\setlength{\parskip}{0pt}
\setlength{\parsep}{0pt}
\item[7.] Analytical solutions for $y$-independent shocks
\item[8.] Alternative derivation of the $\delta$ and $\zeta$ solutions
\item[9.] Examples with initial divergence only
  \begin{enumerate}
     \item[9a.] Formation of a triangular wave
     \item[9b.] Formation of an N-wave
  \end{enumerate}
\item[10.] Examples with initial vorticity only
  \begin{enumerate}
     \item[10a.] Formation of a triangular wave
     \item[10b.] Formation of an N-wave
  \end{enumerate}
\item[11.] Concluding remarks
\end{enumerate}
\end{enumerate}

\newpage

\centerline{\bf I. Advection Equation and Burgers' Equation}

\section{Introduction}                         

\begin{figure*}[t]                                  
\centerline{\includegraphics[width=6.5in]{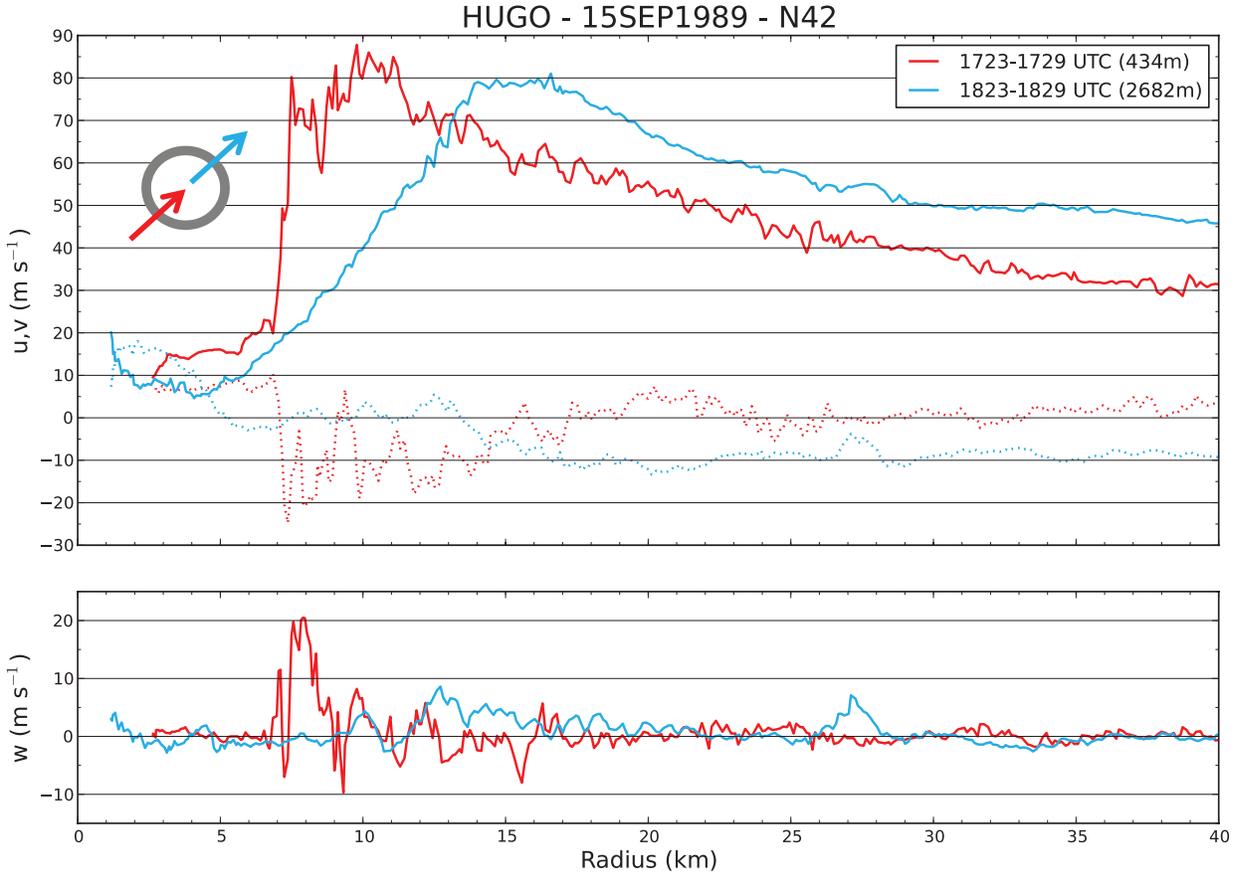}}
\caption{NOAA WP-3D (N42RF) aircraft data from $\sim$400 m (red, inbound,
southwest quadrant) and $\sim$2700 m (blue, outbound, northeast quadrant)
flight legs in Hurricane Hugo on 15 September 1989. In the upper panel
the solid curves show the tangential wind component
while the dotted curves show the radial wind component. The lower
panel shows the vertical component of the velocity. These radial
profiles are based on 1 second flight data, which corresponds to a spatial
resolution of approximately 100 m. Flight data courtesy of NOAA/HRD. From \citet{williams13}.}
\end{figure*}

     The NOAA WP-3D aircraft data obtained in Hurricane Hugo (1989) alerted
the tropical cyclone research community to the dangers of the tropical cyclone
boundary layer and led to research into the possibility that discontinuities
(or shocks) in the boundary layer radial and tangential flow can occur in intense
hurricanes. This data, which has been discussed in detail by \citet{marks08}, is
reproduced here as Fig.~1. As the aircraft flew at $z\approx400$ m northeastward
towards the eye, the boundary layer tangential wind (solid red curve in the
upper panel) increased from 50 m s$^{-1}$ near $r=22$ km to a maximum of
88 m s$^{-1}$ near $r=10$ km.
At the inner edge of the eyewall, there were multiple updraft-downdraft couplets
(the strongest updraft just exceeding 20 m s$^{-1}$) with associated oscillations
of the boundary layer radial and tangential velocity components and a very rapid
60 m s$^{-1}$ change in tangential velocity near $r=7$ km. After ascending in
the eye, the aircraft departed the storm at $z\approx2700$ m (i.e., above the frictional
boundary layer), obtaining the horizontal and vertical velocity data shown by
the blue curves in Fig.~1. If the tangential wind at $z\approx2700$ m is assumed to
be close to gradient balance and the pressure gradient in the boundary layer
is essentially the same as that at $z\approx2700$ m, then the region $r<13$ km has
supergradient boundary layer flow, while the region $r>13$ km has subgradient
boundary layer flow. This is a telltale sign of the importance of the nonlinear
advective effects that produce the near discontinuities in $u$ and $v$ at
$r \approx 7$ km and the near singularity in $w$ at $r \approx 8$ km.

     After the Hugo flight, the risks involved in boundary layer penetrations
into the core of intense hurricanes became more fully appreciated, causing NOAA
to effectively abandon such penetrations after 1989. However, flights above
the boundary layer continued to expand our knowledge of the wind and thermal
structure of intense hurricanes. For example, Fig.~2 shows NOAA WP-3D aircraft
observations of radar reflectivity and radial profiles of tangential wind,
temperature, and dewpoint temperature for Hurricane Frances during a
$3\frac{1}{2}$ hour interval on 30 August 2004, when the storm was
passing just north of the Virgin Islands. This hurricane, described in detail
by \citet{rozoff08}, originated as an African easterly wave that, on 28 August,
developed into a major hurricane with a minimum sea-level pressure of 948 hPa and
a maximum wind speed of 60 m s$^{-1}$. During the time interval shown in
Fig.~2, Frances had well-defined concentric eyewalls, with a 30 km diameter
inner eyewall and a 100 km diameter
outer eyewall. Temperatures near the center were as much as $10^\circ$C warmer
than at  radii of 60--70 km, with Fig.~2f showing a warm-ring structure just
inside the inner eyewall.  In the subsiding air of the echo-free moat between the
concentric eyewalls, dewpoint depressions as large as $6^\circ$C were observed.
Understanding the formation and evolution of such concentric eyewalls is presently
an area of active research, with the boundary layer playing an important role
in the organization of the moist convection.

\begin{figure*}[!th]                                  
\centerline{\includegraphics[width=39pc]{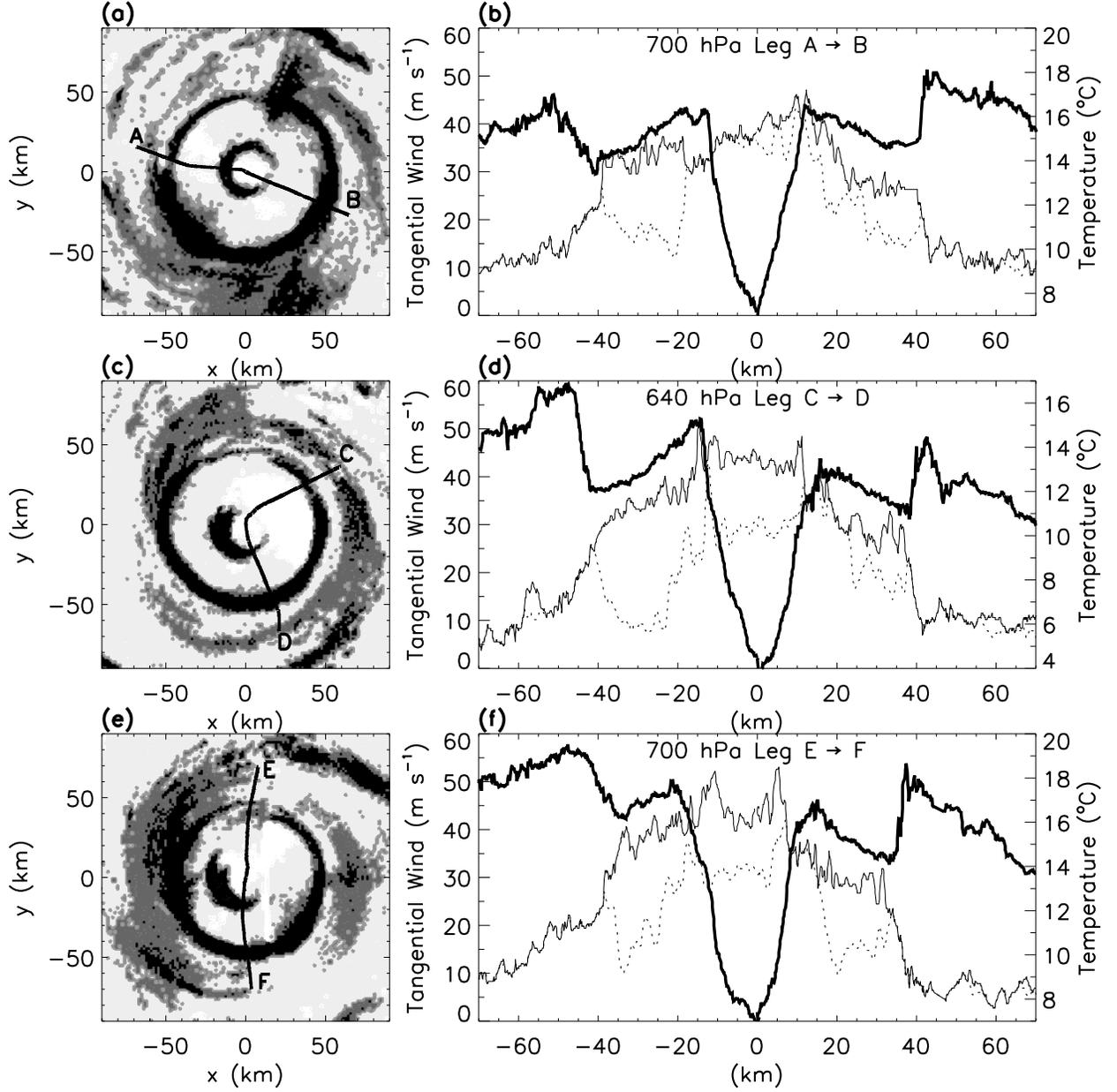}}
\caption{Radar reflectivity and radial profiles of flight-level tangential
wind (m s$^{-1}$; thick solid), temperature ($^\circ$C; thin solid), and dewpoint
temperature ($^\circ$C; dashed) for Hurricane Frances from
1800--1826 UTC (Leg A$\to$B), 1919--1947 UTC (Leg C$\to$D), and
2104--2129 UTC (Leg E$\to$F) on 30 August 2004. Note the large dewpoint
depressions in the moat between the concentric eyewalls. An inner core warm ring
thermal structure is particularly evident on Leg E$\to$F. From \citet{rozoff08}.}
\end{figure*}

\begin{figure}[t]                         
\centerline{\includegraphics[width=22pc]{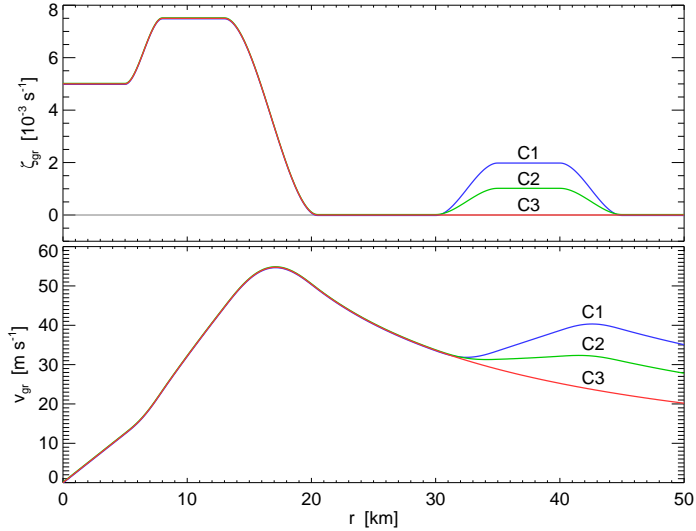}}
\caption{Radial distribution of the forcing $v_{\rm gr}(r)$ (bottom panel)
and the associated vorticity $\zeta_{\rm gr}(r)$ (top panel) for cases C1, C2,
and C3 of the numerical model. All three forcing profiles have the same
$\zeta_{\rm gr}(r)$ and the same $v_{\rm gr}(r)$ for $r \le 30$ km. From \citet{slocum14}.}
\end{figure}

\begin{figure}[b!t]                  
\centerline{\includegraphics[width=22pc]{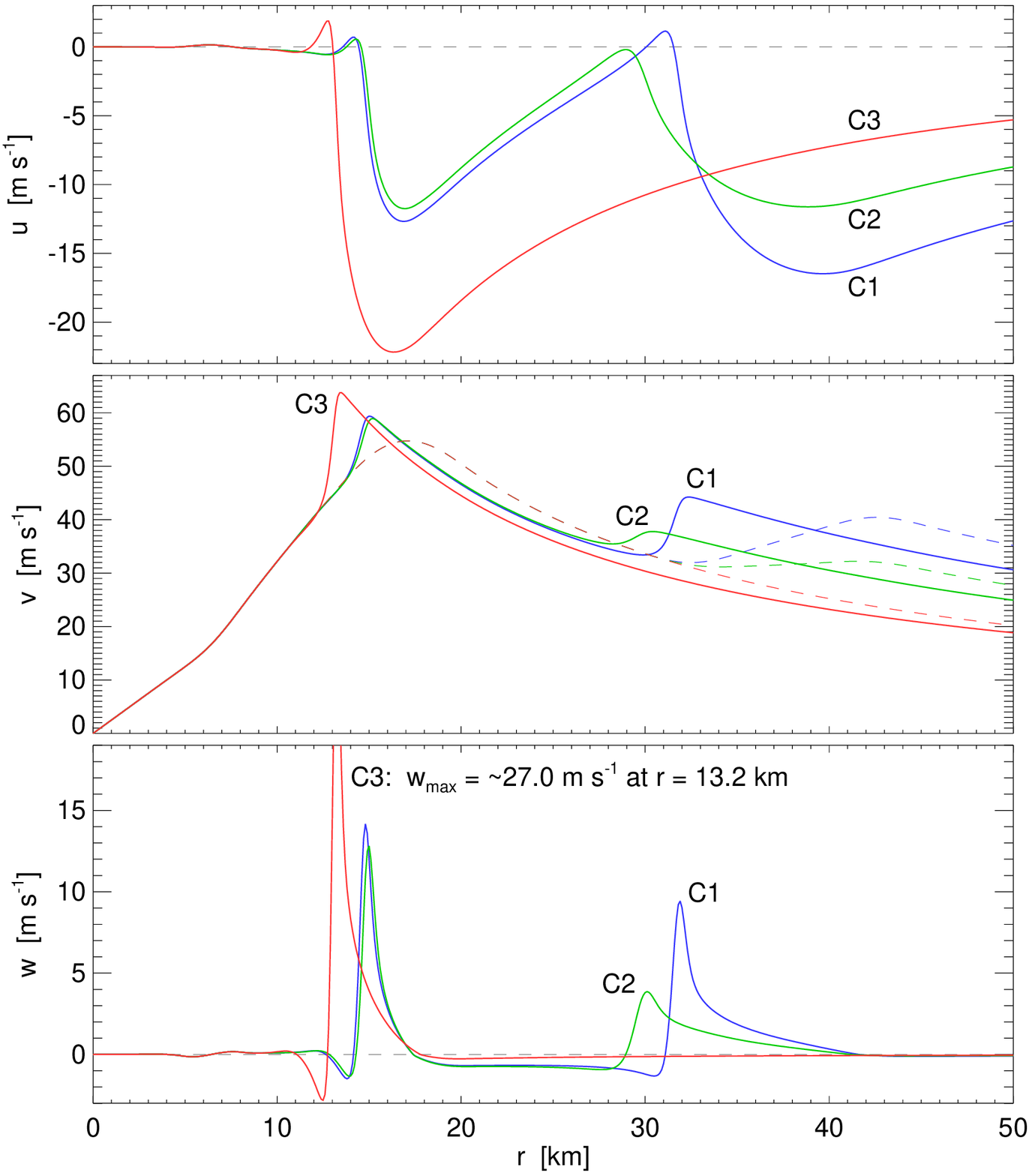}}
\caption{Steady-state slab boundary layer radial profiles of radial velocity
$u$ (top panel), tangential velocity $v$ (middle panel), and vertical velocity
$w$ (bottom panel), for the three forcing profiles shown in Fig.~3. The
radial profile of $w$ for the case with no concentric eyewall reaches a peak
of 27 m~s$^{-1}$, but has been cut off at 19 m~s$^{-1}$ for clarity of the
other profiles. The slab boundary layer model has been solved on the domain
$0 \le r \le 1000$ km, but only the region $0 \le r \le 50$ km is displayed.
In the following sections the radial inflow for case C3 will be interpreted
as a triangular wave while the radial inflows for cases C1 and C2 will be
interpreted as N-waves. From \citet{slocum14}.}
\end{figure}

     We shall argue here that the remarkable convective organization in
hurricanes like Hugo and Frances is primarily due to boundary
layer dynamics, in particular to the formation of discontinuities in the
boundary layer radial inflow and hence singularities in the boundary layer
pumping. In terms of the axisymmetric form of the slab boundary layer
approximation, tropical cyclone boundary layer dynamics can be described by
\begin{equation}                                  
  \begin{split}
     \frac{\partial u}{\partial t} + u\frac{\partial u}{\partial r}
      - \left(f + \frac{v}{r}\right)v + \frac{\cD U}{h}u
     &= K\frac{\partial}{\partial r}\left(\frac{\partial(ru)}{r\partial r}\right)
      - \frac{1}{\rho}\frac{\partial p}{\partial r},  \\
     \frac{\partial v}{\partial t} + u\frac{\partial v}{\partial r}
     + \left(f + \frac{v}{r}\right)u + \frac{\cD U}{h}v
     &= K\frac{\partial}{\partial r}\left(\frac{\partial(rv)}{r\partial r}\right),
  \end{split}
\label{eq1.1}
\end{equation}
where $u$ is the radial component, $v$ the tangential component,
$U=(u^2+v^2)^{1/2}$ the wind speed, and where the Coriolis parameter $f$,
the boundary layer depth $h$, the drag coefficient $\cD$, and the horizontal
diffusivity $K$ are assumed to be constants. The specified forcing term
$-(1/\rho)(\partial p/\partial r)$ can also be interpreted as a specified
gradient wind, since the gradient wind $v_{\rm gr}$ is defined in terms
of the boundary layer density and pressure by $(f+v_{\rm gr}/r)v_{\rm gr}
=(1/\rho)(\partial p/\partial r)$.  \citet{slocum14} presented three
numerical experiments with a slightly generalized version of the slab
boundary layer equations (\ref{eq1.1}). Their generalized version includes
vertical advection terms and an empirical relation for $\cD$ as a function
of $U$. Their specified forcing $v_{\rm gr}(r)$ is shown in the lower panel
of Fig.~3, with the associated relative vorticity $\zeta_{\rm gr}(r)$ shown
in the upper panel.  All three forcing profiles
have the same $v_{\rm gr}(r)$ and the same $\zeta_{\rm gr}(r)$ for
$r \le 30$ km. For experiments C1 and C2, the $\zeta_{\rm gr}(r)$ profiles
have been locally ($30 < r < 45$ km) enhanced over that of experiment C3
so that the associated $v_{\rm gr}(r)$ profiles differ for $r > 30$ km.
The sequence C3$\to$C2$\to$C1 can be considered as an enhancement of the
outer gradient balanced flow while the inner core balanced flow remains
unchanged. For each of these three specified $v_{\rm gr}(r)$ forcing
functions, the numerical model was integrated until a steady state was
obtained. Such steady states are generally obtained quickly with most
of the change from the initial conditions
$u(r,0)=0$ and $v(r,0)=v_{\rm gr}(r)$ occurring in the first hour and only
small changes occurring after 3 hours. Figure 4 shows the steady-state
boundary layer flows beneath each of these three forcing functions.  The three
panels show radial profiles ($0 \le r \le 50$ km) of the
boundary layer radial wind $u$ (top panel), tangential wind $v$ (middle panel),
and vertical velocity $w$ (bottom panel). Note that in each case, strong
radial inflow, supergradient or subgradient tangential winds, and large boundary
layer pumping develop. Due to the $u(\partial u/\partial r)$ term in the
radial equation of motion, Burgers' shock-like structures develop just
inside the local maxima in the initial tangential wind.  At the inner
eyewall ($r \approx 16.5$ km), the maximum radial inflows are 22 m~s$^{-1}$
for case C3, 11.5 m~s$^{-1}$ for case C2, and 12.5 m~s$^{-1}$ for case C1,
so the strength of the inner eyewall shock is considerably reduced by the
presence of an outer shock. Note that, even though cases C1 and C2 have
stronger inflow than case C3 at $r \approx 40$ km, the situation is reversed
at $r \approx 30$ km, a radius at which the radial inflow has been reduced
to essentially zero for cases C1 and C2. Although the radial inflows for
cases C1 and C2 do somewhat recover in the moat region between the two eyewalls
($16.5 < r < 29$ km), the width of the moat and the strength of the agradient term
$[f + (v+v_{\rm gr})/r](v-v_{\rm gr})$ are not large enough to allow a full
recovery of the radial inflow, leading to an inner eyewall boundary layer pumping
(bottom panel of Fig.~4) that is reduced to approximately 50\% of the value
obtained in case C3. In sections 2--5, the general structure of the radial
flow and boundary layer pumping will be related to simple solutions of
the nonlinear advection equation and Burgers' equation.
In sections 2 and 4, it will be shown that the radial inflow
in case C3 resembles an asymptotic triangular wave, while in sections 3 and 5
it will be shown that the radial inflows in cases C1 and C2 resemble an asymptotic
N-wave.\footnote{Although the term ``inverted N-wave" may be more precise, we
use the generic term ``N-wave" throughout the discussion here.}

    This paper is organized into two parts. Part I discusses
analytical solutions of the nonlinear advection equation for asymptotic triangular
waves (section 2) and asymptotic N-waves (section 3).  These two sections review
the concepts of hyperbolic equations,
the method of characteristics, expansive and compressive
regions, wave breaking, multivalued solutions, and the introduction of shock
conditions to guarantee single-valued solutions.  Sections 4 and 5 discuss the
analogous solutions for Burgers' equation that can be solved analytically
via the Cole--Hopf transformation.  Since Burgers' equation includes
the horizontal diffusion term, multivalued solutions do not arise, so shock
conditions are not required. However, for small values of the diffusion
coefficient, the asymptotic triangular wave and the asymptotic N-wave
closely resemble those for the advection equation. Sections 2--5 treat
line-symmetric problems in the Cartesian coordinate and might be called
``toy models" or ``metaphors" for certain aspects of tropical cyclone boundary
layer dynamics. They are presented here to help understand the boundary layer
inflow features that are associated with the advection and diffusion terms
in (\ref{eq1.1}).  In section 6, we
consider analytical solutions of Burgers' equation for the case of
circular symmetry. This axisymmetric case provides further insight into
the formation, propagation, and merger of tropical cyclone boundary layer
shocks. Part II (sections 7--10) discusses analytical solutions of the
line-symmetric version of (\ref{eq1.1}), thus illustrating how multivalued
boundary layer solutions can appear in finite time and how the singularities
can be either divergence-preferred or vorticity-preferred. The analytical solutions
are used to better understand the role of boundary layer shocks in tropical
cyclone dynamics. Section 11 presents some concluding remarks, including the
implications of the present work on understanding eyewall replacement cycles.

\section{Asymptotic triangular waves and their conceptual
         connection with primary eyewalls}       

     We begin our analysis with the one-dimensional, nonlinear advection problem
\begin{equation}                                 
     \frac{\partial u}{\partial t} + u\frac{\partial u}{\partial x} = 0,
     \,\,\,\text{with}\,\,\,   u(x,0) = u_0(x),
\label{eq2.1}
\end{equation}
where the initial condition $u_0(x)$ is a specified function.
In this section, we assume that $u_0(x)$ has the constant value $U$
for $x\leq -a$ and $x\geq0$, and has values $u_0(x)<U$ for $-a<x<0$. Our
example assumes $U<0$, so we are envisioning a boundary layer inflow
toward a cyclone center that lies to the left.
In our discussion of the asymptotic behavior of the solutions of
(\ref{eq2.1}), we shall not be concerned with the details of $u_0(x)$
in the region $-a<x<0$, but rather only with the fact that $u_0(x)<U$
in this region.  As will be seen, the details of $u_0(x)$ are
forgotten as the solution evolves and only the constant $U$ and
the initial integrated momentum anomaly $M=\int_{-a}^0 [U-u_0(x)]dx$
are remembered at large times.

    In order to anticipate some of the discussion to follow, it is interesting
to note that $u=x/(t-t_0)$ is a solution of the nonlinear advection equation with
$t_0$ denoting a positive constant. When $t<t_0$, we have $(\partial u/\partial x)<0$
and the $u(x,t)$ field is steepening with time, i.e.,
$(\partial u/\partial x) \to -\infty$ as $t \to t_0$. In contrast, when
$t>t_0$, we have $(\partial u/\partial x)>0$ and the $u(x,t)$ field is flattening
with time, i.e., $(\partial u/\partial x) \to 0$ as $t \to \infty$.
As we shall see below, we need to fit together these two types of solutions
and ensure that the result is not multivalued. This gives rise to the concepts
of asymptotic triangular waves (this section) and asymptotic N-waves (next section).

     Problem (\ref{eq2.1}) is a hyperbolic equation that can also
be stated in the characteristic form
\begin{equation}                                 
     \frac{du}{dt} = 0  \quad \text{on} \quad \frac{dx}{dt} = u,
\label{eq2.2}
\end{equation}
where $(d/dt)=(\partial/\partial t)+u(\partial/\partial x)$ is the
derivative along a characteristic. Since it follows from \eqref{eq2.2}
that $u$ is invariant along a characteristic and that the characteristics
are therefore straight lines in the $(x,t)$-plane, the solution is
\begin{equation}                                 
     u(x,t) = u_0(\hat{x}),  \quad\text{with}\quad
          x = \hat{x} + u_0(\hat{x}) t,
\label{eq2.3}
\end{equation}
where $\hat{x}(x,t)$ is the label (i.e., the initial position) of the
characteristic that goes through the point $(x,t)$. The continuous
solution (\ref{eq2.3}) is valid only until the shock formation time,
after which the discontinuity in the solution needs to be tracked via a
shock-fitting procedure.

\begin{figure}[!t]             
\centerline{\includegraphics[width=19pc]{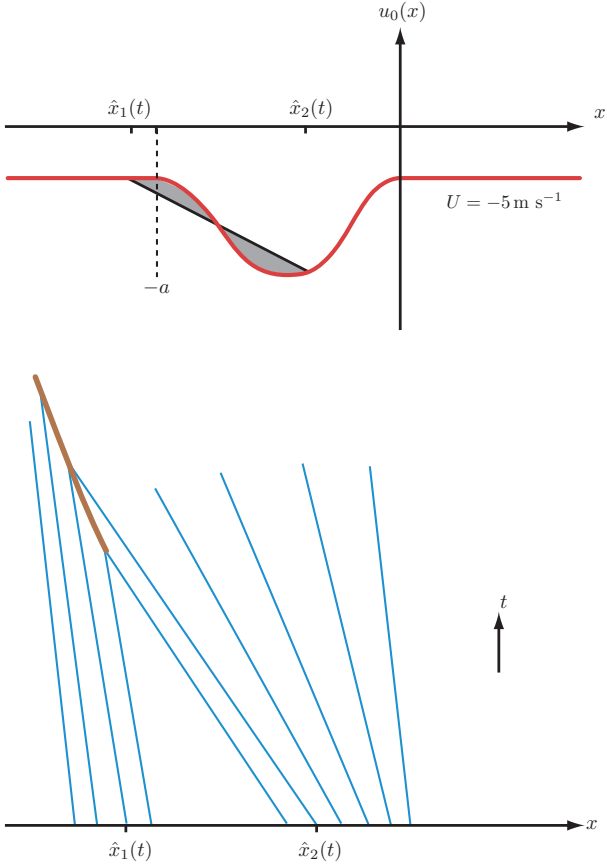}}
\caption{The upper panel shows the initial condition $u_0(x)$, which
has the constant value $U=-5$ m s$^{-1}$ for $x\leq-a$ and $x\geq0$, and
has enhanced inflow, $u_0(x)<U$, in the region $-a<x<0$. The lower
panel shows selected characteristics in the $(x,t)$-plane. According
to the equal area property, the two characteristics labeled $\hat{x}_1(t)$
and $\hat{x}_2(t)$ will simultaneously reach the shock at time $t$ if the
secant line between them cuts off equal areas of the $u_0(x)$ curve,
as indicated by the gray shaded region in the upper panel.}
\end{figure}

    If the shock position at time $t$ is denoted by $x_s(t)$, then from
the second part of (\ref{eq2.3}) we obtain
\begin{equation}                                 
  \begin{split}
     x_s(t) &= \hat{x}_1(t) + u_0(\hat{x}_1(t)) t \,\,\text{and}  \\
     x_s(t) &= \hat{x}_2(t) + u_0(\hat{x}_2(t)) t,
  \end{split}
\label{eq2.4}
\end{equation}
where $\hat{x}_1(t)$ and $\hat{x}_2(t)$ are the values of $\hat{x}$ on
either side of the shock at time $t$.  Elimination of $x_s(t)$ between
the two equations in (\ref{eq2.4}) yields
\begin{equation}                                 
        \hat{x}_2(t) - \hat{x}_1(t)
      = \left[u_0(\hat{x}_1(t)) - u_0(\hat{x}_2(t))\right]t.
\label{eq2.5}
\end{equation}
This is one relation between $\hat{x}_1(t)$, $\hat{x}_2(t)$, and the
specified initial condition $u_0(x)$. A second relation can be found
from Whitham's equal area property, which can be illustrated as
follows \citep{whitham74}. For the given
time $t$, place the points $\hat{x}_1(t)$ and $\hat{x}_2(t)$ on the
$u_0(x)$ curve shown in Fig.~5. According to the equal area property,
these two points will simultaneously reach the shock at time $t$ if the
secant line between them cuts off equal areas of the $u_0(x)$ curve.
This equal area property can be expressed as
\begin{equation}                                 
  \begin{split}
     &\tfrac{1}{2}\left[2U - u_0(\hat{x}_1(t)) - u_0(\hat{x}_2(t))\right]
                 \left[\hat{x}_2(t) - \hat{x}_1(t)\right]   \\
     &\qquad\qquad = \int_{\hat{x}_1(t)}^{\hat{x}_2(t)} [U - u_0(x)]\,dx.
  \end{split}
\label{eq2.6}
\end{equation}
As time proceeds, $\hat{x}_1(t)$ decreases and eventually becomes less than $-a$,
after which $u_0(\hat{x}_1(t))=U$ and the lower limit of the integral in
(\ref{eq2.6}) can be set to $-a$. Equations (\ref{eq2.5}) and (\ref{eq2.6})
then simplify to
\begin{equation}                                 
      \hat{x}_2(t) - \hat{x}_1(t) = \left[U - u_0(\hat{x}_2(t))\right] t,
\label{eq2.7}
\end{equation}
\begin{equation}                                 
     \tfrac{1}{2}\left[U - u_0(\hat{x}_2(t))\right]\left[\hat{x}_2(t) - \hat{x}_1(t)\right]
     = \int_{-a}^{\hat{x}_2(t)} [U - u_0(x)]\,d x.
\label{eq2.8}
\end{equation}
Eliminating $\hat{x}_2(t) - \hat{x}_1(t)$ between these last two equations, we
obtain
\begin{equation}                                 
     \tfrac{1}{2}\left[U - u_0(\hat{x}_2(t))\right]^2 t
     = \int_{-a}^{\hat{x}_2(t)} [U - u_0(x)]\,d x.
\label{eq2.9}
\end{equation}
As time proceeds further, $\hat{x}_2(t)$ increases and eventually reaches zero,
after which, equation (\ref{eq2.9}) yields
\begin{equation}                                 
     \tfrac{1}{2}\left[U - u_0(\hat{x}_2(t))\right]^2 t = M,
\label{eq2.10}
\end{equation}
where the initial integrated momentum anomaly is defined by
\begin{equation}                                 
      M = \int_{-a}^0 [U - u_0(x)] dx > 0.
\label{eq2.11}
\end{equation}
From (\ref{eq2.3}) and (\ref{eq2.10}), we obtain the asymptotic formula
\begin{equation}                                 
      u(x_s(t),t) = u_0(\hat{x}_2(t)) \sim U - \sqrt{2M/t}
\label{eq2.12}
\end{equation}
for the value of $u$ just behind (i.e., just to the right of)
the leftward-moving shock.
From (\ref{eq2.4}), the asymptotic form of the shock position is
\begin{equation}                                 
     x_s(t) \sim Ut - \sqrt{2Mt}.
\label{eq2.13}
\end{equation}
Therefore, the asymptotic form of the solution is
\begin{equation}                                 
   u(x,t) \sim
          \begin{cases}
             U    & \text{if } \qquad\quad        -\infty < x < Ut - \sqrt{2Mt} \\
             x/t  & \text{if }            Ut - \sqrt{2Mt} < x \le Ut   \\
	     U    & \text{if } \qquad\qquad          Ut \le x < \infty,
          \end{cases}
\label{eq2.14}
\end{equation}
which is a triangular wave as plotted in Fig.~6.
The jump in $u$ across the shock is $\sqrt{2M/t}$ and the width of
the triangular region behind the shock is $\sqrt{2Mt}$, so the area
under the $U-u$ curve remains equal to its initial value $M$.
Since the asymptotic formula (\ref{eq2.14}) involves
only $U$ and $M$, the details of the initial condition $u_0(x)$ are
lost. The region where $u \ne U$ might be called the ``forgetful region."
To summarize, all initial conditions with the same $U$ and $M$
have the same ultimate behavior. A smooth initial pulse of radial
inflow evolves into an asymptotic triangular wave, with a discontinuity
in the radial velocity and a singularity in the boundary layer pumping.
This is conceptually similar to the $u$ profile of case C3 in the top
panel of Fig.~4.

\begin{figure}[!t]             
\centerline{\includegraphics[width=19pc]{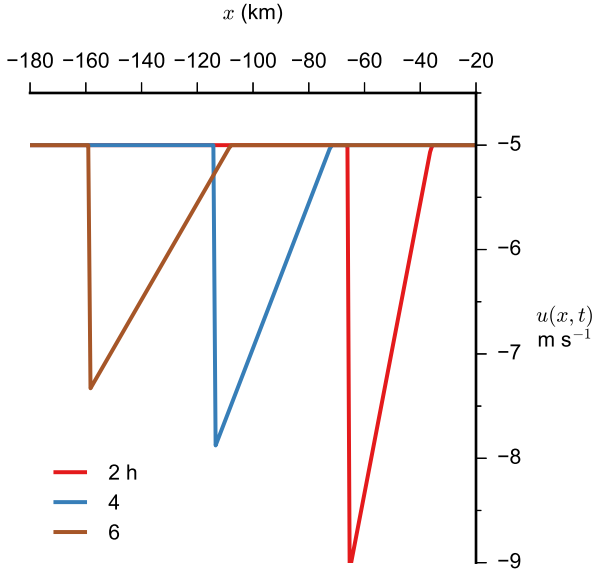}}
\caption{The advection equation asymptotic solution (\ref{eq2.14}),
with $u(x,t)$ plotted as a function of $x$ for $U=-5$ m\, s$^{-1}$,
$M=60,000$ m$^2$~s$^{-1}$, and $t=2,4,6$ h. The strength of the
shock decreases as $t^{-1/2}$, being $4.08,\, 2.89,\,2.36$ m s$^{-1}$
at $t=2,\, 4,\, 6$ h. The width of the triangular region behind
the shock increases as $t^{1/2}$, being $29.4,\, 41.6,\,50.9$ km
at $t=2,\, 4,\, 6$ h. As it moves to the left, the shock slows
down, its velocity being given by $U-\sqrt{2M/t}$, which has the
values $-9.08,\, -7.89,\,-7.36$ m s$^{-1}$ for $t=2,\, 4,\, 6$ h.}
\end{figure}

\section{Asymptotic N-waves and their conceptual connection
         with moats and double eyewalls}           

\subsection{Undamped N-waves}

     In the previous section, we presented some ideas concerning the
question of how a smooth pulse of enhanced radial inflow evolves into
a primary eyewall shock. We now consider the following related question:
How does a smooth undulation of enhanced and reduced radial inflow
evolve into double eyewall shocks? The initial condition for this section is illustrated
in the upper panel of Fig.~7. Since there are two compressive regions where
$(\partial u_0/\partial x)<0$ surrounding a single expansive region where
$(\partial u_0/\partial x)>0$, we expect two shocks to form. The initial
integrated momentum anomalies for the forward and rear areas are defined by
\begin{equation}                                 
  \begin{split}
      M_{\rm f} &= \int_a^b [U - u_0(x)]\, dx > 0\,\text{and} \\
      M_{\rm r} &= \int_b^c [u_0(x) - U]\, dx > 0,
  \end{split}
\label{eq3.1}
\end{equation}
where $M_{\rm f}$ is the left enhanced area and $M_{\rm r}$ is
the right reduced area.
The characteristics for this problem are shown in the lower panel of
Fig.~7 and the asymptotic solution is given by
\begin{equation}                                 
   u(x,t) \sim \begin{cases}
                 U    & \text{if } \qquad\quad     -\infty < x < Ut-\sqrt{2M_{\rm f}t} \\
                 x/t  & \text{if }   Ut-\sqrt{2M_{\rm f}t} < x < Ut+\sqrt{2M_{\rm r}t} \\
	         U    & \text{if }   Ut+\sqrt{2M_{\rm r}t} < x < \infty,
               \end{cases}
\label{eq3.2}
\end{equation}
which is an N-wave as plotted in Fig.~8 for $U=-5$ m\, s$^{-1}$ and $t=2,4,6$ h.
Figure 8a is for the choice $M_{\rm f}=M_{\rm r}=60,000$ m$^2$~s$^{-1}$,
which produces forward and rearward shocks of equal strength. Figure 8b is
for the choice $M_{\rm f}=60,000$ m$^2$~s$^{-1}$ and $M_{\rm r}=30,000$ m$^2$~s$^{-1}$,
which produces a rearward shock that is weaker than the forward shock.
The jump in $u$ across the front shock is $\sqrt{2M_{\rm f}/t}$, while the jump
across the rear shock is $\sqrt{2M_{\rm r}/t}$. The width of the region between
the two shocks is $\sqrt{2M_{\rm f}t}+\sqrt{2M_{\rm r}t}$. The area under the $U-u$
curve in the left portion of the N-wave remains equal to its initial value
of $M_{\rm f}$, while the area under the $U-u$ curve in the right portion of the
N-wave remains equal to its initial value of $M_{\rm r}$.

\begin{figure}[!t]             
\centerline{\includegraphics[width=19pc]{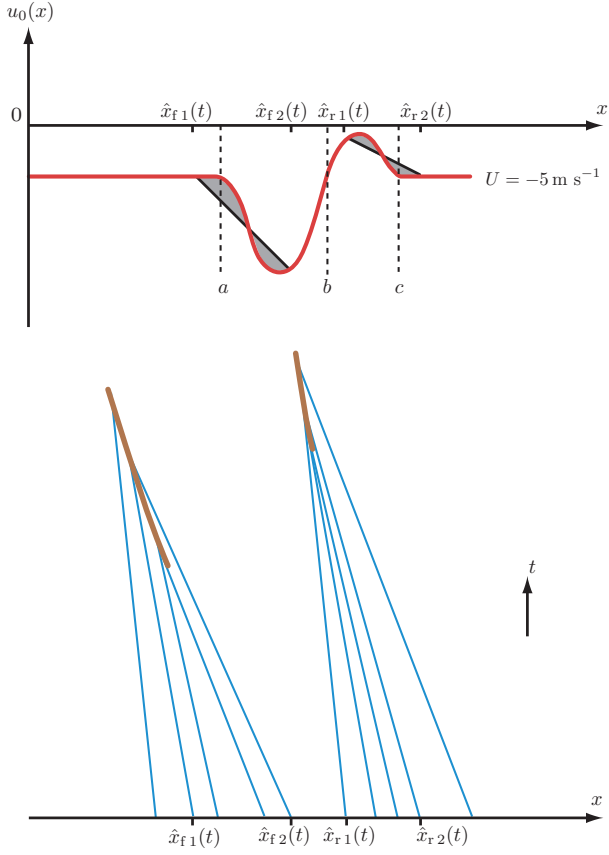}}
\caption{The upper panel shows the initial condition $u_0(x)$, which
has the constant value $U=-5$ m s$^{-1}$ for $x\leq a$, $x\geq c$, and
$x=b$;
enhanced inflow, $u_0(x)<U$, in the region $a<x<b$; and reduced
inflow, $u_0(x)>U$, in the region $b<x<c$. The lower
panel shows selected characteristics in the $(x,t)$-plane. According
to the equal area property, the two characteristics labeled $x_{f\,1}(t)$
and $x_{f\,2}(t)$ will simultaneously reach the forward shock if the
secant line on the left cuts off equal areas of the $u_0(x)$ curve.
Similarly, the two characteristics labeled $x_{r\,1}(t)$ and $x_{r\,2}(t)$
will simultaneously reach the rear shock if the secant line on the right
cuts off equal areas of the $u_0(x)$ curve. In this example, the forward
shock forms sooner and is stronger. Note that the divergent region between
the shocks becomes wider with time.}
\end{figure}

\begin{figure}[!t]             
\centerline{\includegraphics[width=19pc]{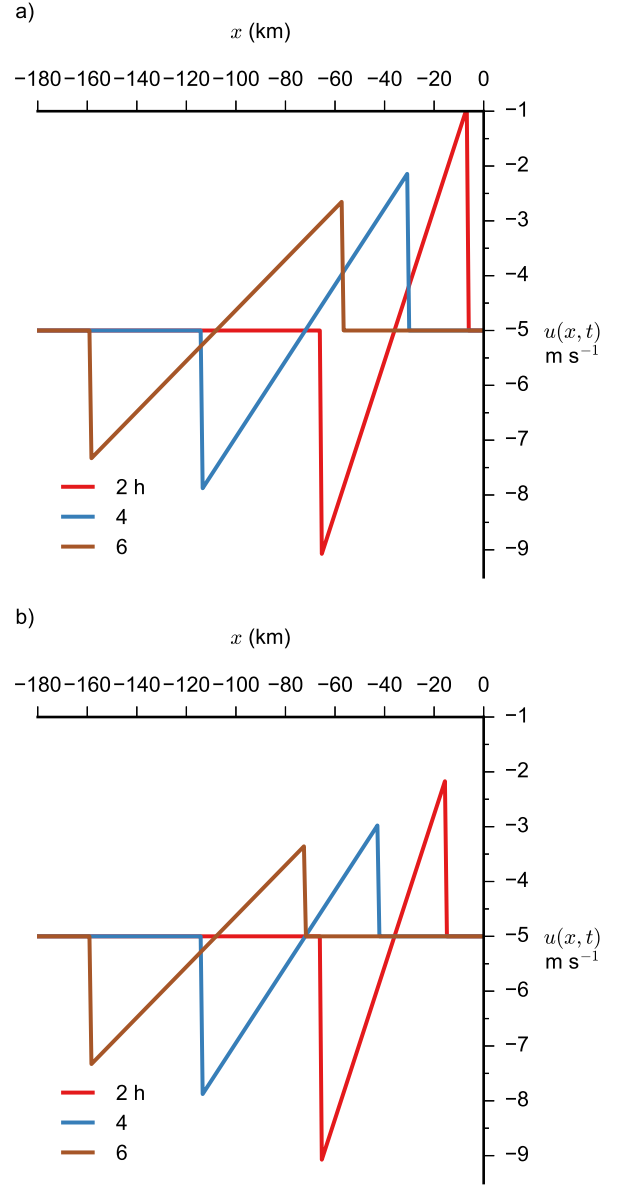}}
\caption{The advection equation asymptotic solution (\ref{eq3.2}),
with $u(x,t)$ plotted as a function of $x$ for $U=-5$ m\, s$^{-1}$
and $t=2,4,6$ h. The top figure is for the choice
$M_{\rm f}=M_{\rm r}=60,000$ m$^2$~s$^{-1}$, which produces forward
and rearward shocks of equal strength. The bottom figure is for the
choice $M_{\rm f}=60,000$ m$^2$~s$^{-1}$ and $M_{\rm r}=30,000$ m$^2$~s$^{-1}$,
which produces a rearward shock that is weaker than the forward shock.}
\end{figure}

This N-wave pattern for the nonlinear advection equation is similar to
the N-wave patterns shown in the top panel of Fig.~4 for the slab boundary
layer model simulations of concentric eyewalls (cases C1 and C2).

\subsection{Damped N-waves}

     This section discusses how initial conditions that result in two
shocks (i.e., N-waves) in the undamped problem (\ref{eq2.1}) can lead
to two, one, or no shocks in the damped problem. We begin the analysis
with the damped nonlinear advection problem
\begin{equation}                                 
        \frac{\partial u}{\partial t}
     + u\frac{\partial u}{\partial x} = -\frac{u}{\tau},
        \,\,\,\text{with}\,\,\,   u(x,0) = u_0(x),
\label{eq3.3}
\end{equation}
where $\tau$ is the constant damping time scale and the initial condition
$u_0(x)$ is a specified function. Problem (\ref{eq3.3}) is a hyperbolic
equation that can also be stated in the characteristic form
\begin{equation}                                 
     \frac{d}{dt}\left(ue^{t/\tau}\right) = 0
              \quad \text{on} \quad \frac{dx}{dt} = u,
\label{eq3.4}
\end{equation}
where $(d/dt)=(\partial/\partial t)+u(\partial/\partial x)$ is the
derivative along a characteristic. Since $ue^{t/\tau}$ is invariant along
each characteristic, the solution of the first equation in (\ref{eq3.4})
is
\begin{equation}                                 
     u(x,t) = u_0(\hat{x})e^{-t/\tau},
\label{eq3.5}
\end{equation}
where $\hat{x}(x,t)$ is the label (i.e., the initial position) of the
characteristic that goes through the point $(x,t)$. Using the solution
(\ref{eq3.5}) in the right-hand side of the second equation in (\ref{eq3.4})
and then integrating in time, we obtain
\begin{equation}                                 
        x = \hat{x} + \hat{t} u_0(\hat{x}),
                     \quad \text{where} \quad
	\hat{t}(t) = \tau\left(1-e^{-t/\tau}\right).
\label{eq3.6}
\end{equation}
The characteristics defined by (\ref{eq3.6}) are not straight lines in the
$(x,t)$-plane, although they do become straight in the limit $\tau\to\infty$,
in which case $\hat{t}(t)\to t$.
If shocks appear, the continuous solution (\ref{eq3.5}) and (\ref{eq3.6})
is valid only until the first shock formation time, after which the
discontinuity in the solution needs to be tracked
 via a shock-fitting procedure. The
vertical motion implied by the $u(x,t)$ solution can be found from
$w=-h(\partial u/\partial x)$, where the boundary layer depth $h$ is
taken as 1000 m. Using (\ref{eq3.5}) and (\ref{eq3.6}), we obtain the
boundary layer pumping formula
\begin{equation}                                 
    w(x,t) = -h\left(\frac{u_0'(\hat{x})}{1+\hat{t}u_0'(\hat{x})}\right)e^{-t/\tau},
\label{eq3.7}
\end{equation}
where $u_0'(\hat{x})$ is the first derivative of the initial condition $u_0(\hat{x})$.
A singularity in $w$ will occur along the characteristic $\hat{x}$ if
and when $1+\hat{t}u_0'(\hat{x})=0$, i.e., at the shock formation time
$t_s$ given implicitly by
\begin{equation}                                 
     \hat{t}(t_s) = \frac{1}{[-u_0'(\hat{x})]_{\rm max}}.
\label{eq3.8}
\end{equation}

\begin{figure}[!t]                       
\centerline{\includegraphics[width=19pc]{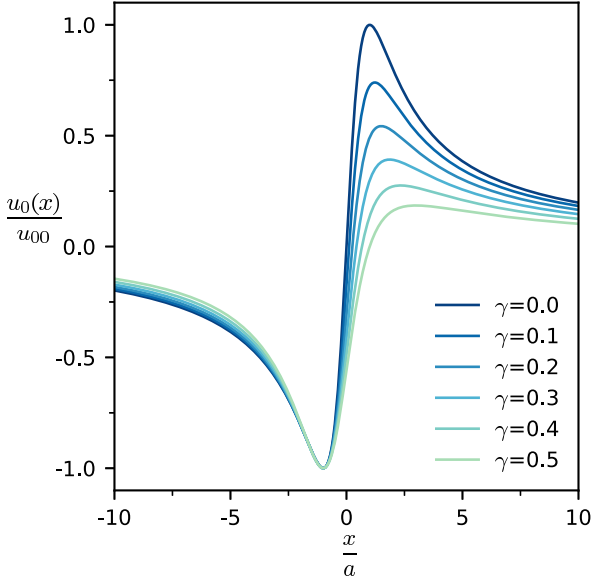}}
\caption{Plots of the initial condition (\ref{eq3.9}) for six different values
of the asymmetry parameter $\gamma$. When $\gamma=0$, the initial condition is
perfectly anti-symmetric about $x=0$.}
\end{figure}

\begin{table}[b!t]                        
\centering
{\begin{tabular}{cccccc}
\hline\hline 
   Case  & $\gamma$ & $\tau$ & $a/\tau u_{00}$ &  $t_{s1}$  &  $t_{s2}$ \\
         &          & (h)    &                 &   (h)      &  (h)  \\
\hline 
    A    &   0.05   & 3.33   &      0.150      &    3.03    &   4.96    \\
    B    &   0.05   & 2.22   &      0.225      &    5.01    & No Shock  \\
    C    &   0.05   & 1.67   &      0.300      &  No Shock  & No Shock  \\
    D    &   0.00   & 3.33   &      0.150      &    3.05    &   3.05    \\
    E    &   0.00   & 1.67   &      0.300      &  No Shock  & No Shock  \\
\hline 
\end{tabular}}
\caption{Data for cases A through E. All cases have an initial horizontal scale of
$a=18$ km and a maximum initial flow of $u_{00}=10$ m s$^{-1}$.  Values of the
initial asymmetry parameter $\gamma$ are given in the second column. Values of the
damping time $\tau$ are given in the third column and the resulting values of
$(a/\tau u_{00})$ are given in the fourth column. The shock formation times
$t_{s1}$ and $t_{s2}$ are given in the last two columns. The values of $t_{s1}$
and $t_{s2}$ have been computed using (\ref{eq3.15}).}
\end{table}

\begin{figure*}[!t]                       
\centerline{\includegraphics[width=39pc]{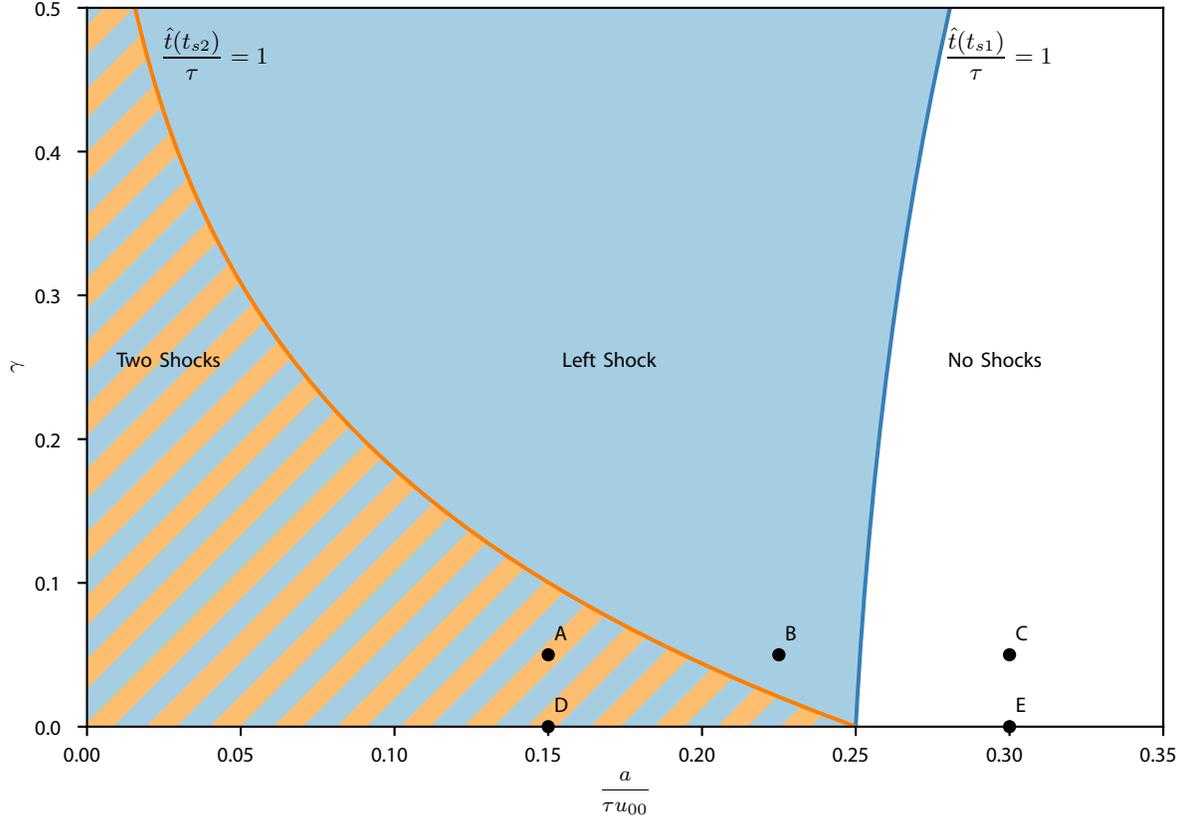}}
\caption{Regions of the $(a/\tau u_{00},\gamma)$-plane where the solutions
(\ref{eq3.5})--(\ref{eq3.6}) contain two shocks, only a left shock, or no shocks.
When the initial condition is perfectly anti-symmetric ($\gamma=0$), the
solutions contain two shocks for $a/\tau u_{00}<0.25$, and no shocks for
$a/\tau u_{00}>0.25$. When the initial disturbance in $u$ is weaker on the
right-hand side (e.g., $\gamma=0.05$), the solutions can have two shocks
for small values of $a/\tau u_{00}$, a single shock on the left-hand side
for intermediate values of $a/\tau u_{00}$, or no shocks for large values
of $a/\tau u_{00}$. The curve separating the hatched region from the blue
region corresponds to $\hat{t}(t_{s2})=\tau$, while the curve separating the
blue region from the white region corresponds to $\hat{t}(t_{s1})=\tau$.
The three dots along $\gamma=0.05$ correspond to the solutions shown in
panels A, B, and C of Fig.~11, while the two dots along $\gamma=0$ correspond
to the solutions shown in panels D and E of Fig.~11.}
\end{figure*}

     As an example, consider the initial condition
\begin{equation}                                 
   \frac{u_0(x)}{u_{00}} = \frac{(2+\gamma)[(1-\gamma)(x/a)-\gamma]}
	                        {(x/a)^2+\gamma(1+\gamma)(x/a)+(1+\gamma)^2},
\label{eq3.9}
\end{equation}
where the initial maximum flow $u_{00}$, the horizontal scale $a$, and
the asymmetry parameter $\gamma$ are specified constants. Plots of
(\ref{eq3.9}) for $\gamma=0.0,\, 0.1,\, 0.2,\, 0.3,\, 0.4,\, 0.5$ are
shown in Fig.~9. Note that, when $\gamma=0$, the $u_0(x)$ field is perfectly
anti-symmetric about $x=0$. Since the derivative of (\ref{eq3.9}) is
\begin{equation}                                 
   \frac{u_0'(x)}{u_{00}/a} = \frac{(2+\gamma)[1+\gamma-(1-\gamma)(x/a)][(x/a)+1]}
	                           {[(x/a)^2+\gamma(1+\gamma)(x/a)+(1+\gamma)^2]^2},
\label{eq3.10}
\end{equation}
it is easily seen that the minimum value of $u_0(x)$ occurs at
$(x/a)=-1$ and the maximum value occurs when $(x/a)=(1+\gamma)/(1-\gamma)$,
with the corresponding values of $u_0(x)$ being $-u_{00}$ and
$[(2+\gamma)/(2-\gamma)][(1-\gamma)/(1+\gamma)]^2 u_{00}$, respectively.
The second derivative of (\ref{eq3.9}) is
\begin{equation}                                 
   \frac{u_0''(x)}{u_{00}/a^2}
           = \frac{2(2+\gamma)(x/a)[(1-\gamma)(x/a)^2-3\gamma (x/a)-3(1+\gamma)]}
                  {[(x/a)^2+\gamma(1+\gamma)(x/a)+(1+\gamma)^2]^3}.
\label{eq3.11}
\end{equation}
From the numerator on the right-hand side of (\ref{eq3.11}), we
see that $u_0''(x)=0$ at $x=0$ and at the two points that are
solutions of the quadratic equation $(1-\gamma)(x/a)^2-3\gamma (x/a)-3(1+\gamma)=0$.
These two solutions, denoted by $x_1/a$ and $x_2/a$, are
\begin{equation}                                 
      \frac{x_{1,2}}{a} = \frac{3\gamma\mp\sqrt{3(4-\gamma^2)}}{2(1-\gamma)}.
\label{eq3.12}
\end{equation}
The points $x_1$ and $x_2$ correspond
to local minima of $u_0'(x)$, while the point $x=0$ corresponds to a
local maximum of $u_0'(x)$.  A shock cannot occur along the
characteristic $\hat{x}=0$ because $u_0'(0)>0$ and $1+\hat{t}u_0'(0)=0$
cannot ever be satisfied. However, shocks can occur along the characteristics
$\hat{x}=x_1$ and $\hat{x}=x_2$. We denote the shock formation time along
these two characteristics as $t_{s1}$ and $t_{s2}$. From (\ref{eq3.8})
and (\ref{eq3.10}), we then obtain
\begin{equation}                                 
     \hat{t}(t_{sj}) = F_j(\gamma) \frac{a}{u_{00}},
\label{eq3.13}
\end{equation}
where
\begin{equation}                                 
    F_j(\gamma) =-\frac{[(x_j/a)^2+\gamma(1+\gamma)(x_j/a)+(1+\gamma)^2]^2}
                       {(2+\gamma)[1+\gamma-(1-\gamma)(x_j/a)][(x_j/a)+1]}
\label{eq3.14}
\end{equation}
for $j=1,2$. Solving (\ref{eq3.13}) for $t_{sj}$, we obtain
\begin{equation}                                 
     t_{sj} = -\tau\ln\left[1 - F_j(\gamma) \frac{a}{\tau u_{00}}\right].
\label{eq3.15}
\end{equation}
This formula has been used to construct Fig.~10, which divides the dimensionless
$(a/\tau u_{00},\gamma)$-plane into three regions. In the hatched region, two
shocks occur since the argument of the natural logarithm in (\ref{eq3.15}) is
positive for both $j=1$ and $j=2$. The shock formation time for the left shock
occurs before the right shock for initial conditions where $\gamma>0$. In the
blue region, a shock occurs only on
the left side since the argument of the natural logarithm is positive only for
$j=1$. In the white region, no shocks occur since the argument of the natural
logarithm is negative for both $j=1$ and $j=2$.  Table 1 lists data for the five
examples indicated by the dots A--E in Fig.~10. The solutions $u(x,t)$ at three
different times are plotted\footnote{In order to avoid iterative procedures in
dealing with the implicit nature of the solutions (\ref{eq3.5})--(\ref{eq3.6}),
a simple way to produce plots of these solutions is as follows. Choose a time
$t$ and then calculate the corresponding $\hat{t}$ from the second entry in
(\ref{eq3.6}). Choose a set of equally spaced values of $\hat{x}$ and then use
the first entry in (\ref{eq3.6}) to calculate the corresponding set of unequally
spaced values of $x$. Then use (\ref{eq3.5}) to calculate $u(x,t)$ at the unequally
spaced $x$-points. Finally, plot $u(x,t)$ as a function of $x$ at the chosen time
$t$ using a plotting routine that can handle unequally spaced data points.}
in Fig.~11. For the cases that produce one or two
shocks (cases A,B,D), the final time is the shock formation time for the left
shock (between 3 and 5 hours, as listed in Table 1). For the cases that don't
produce a shock
(cases C and E), the times are 0, 3, and 6 h. Note that all cases are
characterized by a broadening divergent region with collapsing convergent
regions on each side. Cases A and D have weak damping ($\tau=3.33$ h)
and produce two shocks, while cases C and E have strong damping ($\tau=1.67$ h)
and do not produce shocks. Case B has an intermediate value of damping
($\tau=2.22$ h) and produces a shock only on the left side. Another view of
cases A and C is provided by the characteristic curves shown in Fig.~12. The
upper panel (case A) illustrates the intersection of characteristics on the
left side near $t=3$ h. In the lower panel (case C), the damping is strong enough
that no shocks are produced, even though there is some concentration of
convergence on the left side.

\begin{figure*}[!th]                       
\centerline{\includegraphics[width=39pc]{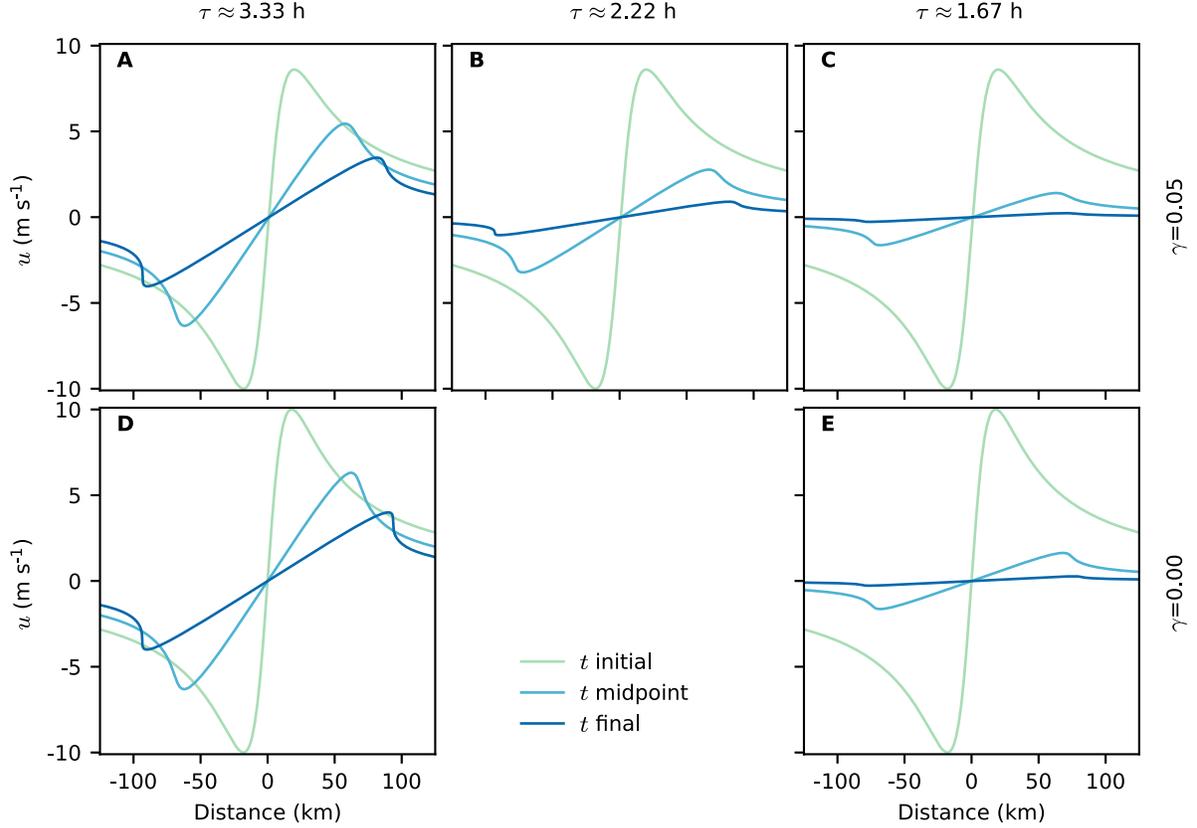}}
\caption{Five sample solutions corresponding to the points A--E of Fig.~10
for the initial condition (green curves), the midpoint of the solution (cyan curves),
and the final time (blue curves). The final time for cases A, B, and D
represent the time of the first shock formation.
For cases D and E, the initial disturbance is perfectly anti-symmetric
($\gamma=0$), while for cases A--C the initial disturbance is stronger
on the left side ($\gamma=0.05$). Damping is strongest for C and E
($\tau=1.67$ h), in which case no shocks form. Damping is weakest for
cases A and D ($\tau=3.33$ h), in which case shocks form on both sides.
For an intermediate value of damping ($\tau=2.22$ h), case B produces
a weak shock on the left side.}
\end{figure*}

\begin{figure}[!t]                       
\centerline{\includegraphics[width=19pc]{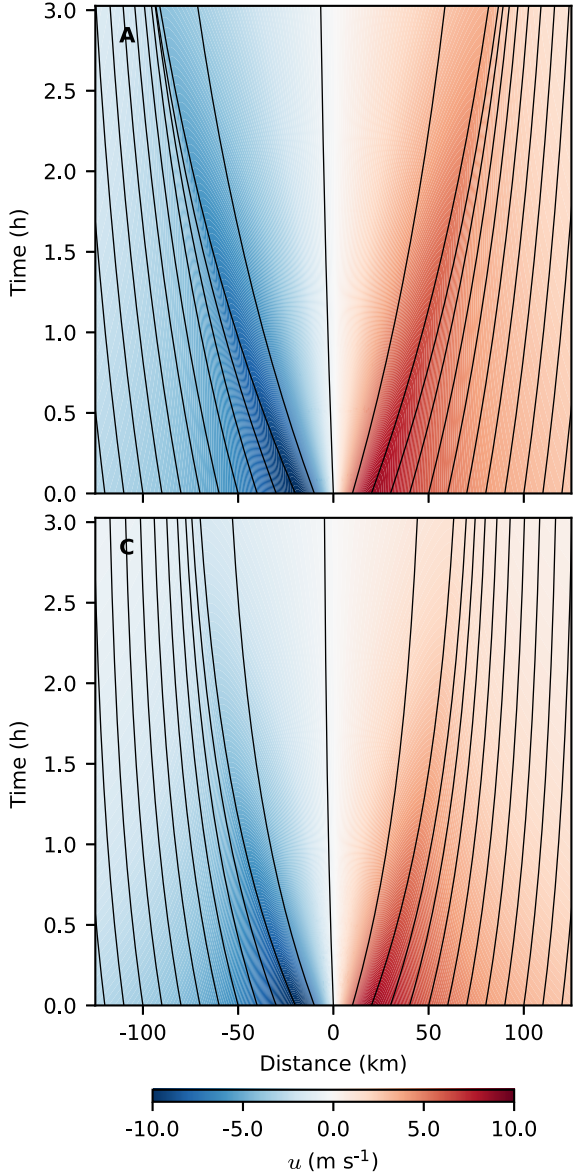}}
\caption{Characteristic curves (solid lines) and $u(x,t)$ (color shading)
for cases A and C of Figs.~10 and 11. Case A produces a shock on the left
side at $t=t_{s1}=3.03$ h, as indicated by the imminent intersection of the
characteristics near $x=-95$ km. A shock also forms on the right side at a
later time ($t=t_{s2}=4.96$ h). Damping is twice as strong in case C.
Even though there is enhanced convergence on the left side of
the divergent region, no shock forms.}
\end{figure}

\section{Triangular waves and primary eyewalls
	 from Burgers' equation}                  

     To further understand the formation and propagation
of boundary layer shocks, we now consider Burgers' equation
\begin{equation}                                 
        \frac{\partial u}{\partial t}
     + u\frac{\partial u}{\partial x}
     = K\frac{\partial^2 u}{\partial x^2},
\label{eq4.1}
\end{equation}
with the initial and boundary conditions
\begin{equation}                                 
        u(x,0) = u_0(x),  \qquad
        u(x,t) \to U \,\,\, \text{as}\,\,\, x \to \pm\infty,
\label{eq4.2}
\end{equation}
where the function $u_0(x)$ and the constant $U$ are specified. For ease of
physical interpretation, we again assume that $U<0$, i.e., a basic inflow
toward the storm center, which lies far to the left of the origin.
Note that the Burgers' equation (\ref{eq4.1}) captures three
important terms in the radial momentum equation of the slab boundary
layer model (\ref{eq1.1}), albeit in the line-symmetric rather than the
axisymmetric form. An excellent general mathematical discussion of Burgers'
equation can be found in the book by \citet{whitham74}.

     In sections 2 and 3, the discussion concerned solutions of the hyperbolic
problems (\ref{eq2.1}) and (\ref{eq3.3}), so that the method of characteristics
played a central role.
Since Burgers' equation (\ref{eq4.1}) is not hyperbolic, the method
of characteristics is not useful. However, considerable analytical progress can
be made using the Cole--Hopf transformation. The mathematical analysis given here
follows that given by \citet{lighthill56}
in his study of viscosity effects in sound waves of finite amplitude.
Although our application to the radial inflow in the tropical cyclone boundary
layer has nothing to do with compressibility effects and finite amplitude sound
waves, we have adapted Lighthill's mathematical analysis to our problem.
We begin by considering solutions of
Burgers' equation with an initial condition consisting of a localized irregularity
superposed on the constant flow $U$. Define the new dependent variable $\hat{u}=u-U$
and the new independent variables $(\hat{x},\hat{t})=(x-Ut,t)$. Then
\begin{equation}                                 
        \frac{\partial}{\partial t}
      = \frac{\partial}{\partial\hat{t}} - U\frac{\partial}{\partial\hat{x}}, \qquad
        \frac{\partial}{\partial x}
      = \frac{\partial}{\partial\hat{x}},
\label{eq4.3}
\end{equation}
and Burgers' equation becomes
\begin{equation}                                 
              \frac{\partial\hat{u}}{\partial\hat{t}}
     + \hat{u}\frac{\partial\hat{u}}{\partial\hat{x}}
     = K\frac{\partial^2\hat{u}}{\partial\hat{x}^2},
\label{eq4.4}
\end{equation}
with the initial and boundary conditions
\begin{equation}                                 
        \hat{u}(\hat{x},0) = u_0(x) - U,  \qquad
        \hat{u}(\hat{x},\hat{t}) \to 0 \,\,\, \text{as}\,\,\, \hat{x} \to \pm\infty.
\label{eq4.5}
\end{equation}

     An integral relation associated with the problem (\ref{eq4.4})--(\ref{eq4.5})
can be obtained by writing (\ref{eq4.4}) in the form
\begin{equation}                                 
       \frac{\partial\hat{u}}{\partial\hat{t}}
     + \frac{\partial}{\partial\hat{x}}
       \left(\frac{1}{2}\hat{u}^2 - K\frac{\partial\hat{u}}{\partial\hat{x}}\right) = 0,
\label{eq4.6}
\end{equation}
and then integrating over the entire domain to obtain the conservation
relation $(dM/dt)=0$, where
\begin{equation}                                 
       M = \int_{-\infty}^\infty \left[U-u(x,t)\right] \, dx
	 = \int_{-\infty}^\infty \left[U-u_0(x)\right] \, dx > 0,
\label{eq4.7}
\end{equation}
so that the integrated momentum $M$ is an invariant of the problem.

     The problem (\ref{eq4.4})--(\ref{eq4.5}) can be solved analytically
using the Cole--Hopf transformation.  The first step in this transformation
is to use (\ref{eq4.6}) to define the velocity potential $\chi(\hat{x},\hat{t})$
such that
\begin{equation}                                 
       \hat{u} = \frac{\partial\chi}{\partial\hat{x}},   \qquad
     \frac{1}{2}\hat{u}^2 - K\frac{\partial\hat{u}}{\partial\hat{x}}
                    = -\frac{\partial\chi}{\partial\hat{t}}.
\label{eq4.8}
\end{equation}
Combining these last two equations, we obtain
\begin{equation}                                 
                         \frac{\partial\chi}{\partial\hat{t}}
     +  \frac{1}{2}\left(\frac{\partial\chi}{\partial\hat{x}}\right)^2
     = K\frac{\partial^2\chi}{\partial\hat{x}^2}.
\label{eq4.9}
\end{equation}
The second step in the Cole--Hopf transformation is to define the new dependent
variable $\varphi(\hat{x},\hat{t})$ by
\begin{equation}                                 
            \varphi = \exp\left(-\frac{\chi}{2K}\right)
                                 \qquad {\rm or}  \qquad
               \chi = -2K \ln\varphi,
\label{eq4.10}
\end{equation}
from which it follows that
\begin{equation}                                 
         \frac{1}{2}\left(\frac{\partial\chi}{\partial \hat{x}}\right)^2
     - K \frac{\partial^2\chi}{\partial\hat{x^2}}
     = \frac{2K^2}{\varphi} \frac{\partial^2\varphi}{\partial\hat{x}^2}.
\label{eq4.11}
\end{equation}
Using (\ref{eq4.11}) in (\ref{eq4.9}), we obtain
\begin{equation}                                 
        \frac{\partial\varphi}{\partial\hat{t}}
     = K\frac{\partial^2\varphi}{\partial\hat{x}^2},
\label{eq4.12}
\end{equation}
with the initial and boundary conditions
\begin{equation}                                 
  \begin{split}
    &\varphi(\hat{x},0) = \exp\left(-\frac{1}{2K}\int_{\hat{x}}^\infty [U-u_0(x')]\, dx'\right), \\
    &\varphi(\hat{x},\hat{t}) \to e^{-R} \quad \text{as}\,\,\,\, \hat{x} \to -\infty,  \\
    &\varphi(\hat{x},\hat{t}) \to 1      \quad \text{as}\,\,\,\, \hat{x} \to  \infty,
  \end{split}
\label{eq4.13}
\end{equation}
where $R=M/2K$ is the Reynolds' number.
Thus, the Cole--Hopf procedure (\ref{eq4.8})--(\ref{eq4.11}) has transformed
the nonlinear advection-diffusion
equation (\ref{eq4.4}) to the linear diffusion equation (\ref{eq4.12}). If we
can solve the diffusion equation (\ref{eq4.12}) for $\varphi(\hat{x},\hat{t})$,
we can recover the solution of the nonlinear equation (\ref{eq4.4}) from
\begin{equation}                                 
     \hat{u} = -\frac{2K}{\varphi} \frac{\partial\varphi}{\partial\hat{x}}.
\label{eq4.14}
\end{equation}
The challenge now is to find a simple solution of (\ref{eq4.12}) and (\ref{eq4.13})
that translates into a physically interesting solution of (\ref{eq4.1}) and
(\ref{eq4.2}).

     An interesting solution of the diffusion equation (\ref{eq4.12}) is
\begin{equation}                                 
  \begin{split}
     \varphi(\hat{x},\hat{t})
         &= 1 + \tfrac{1}{2}\left(e^{-R} - 1\right)
                        \text{erfc}\left(\frac{\hat{x}}{\sqrt{4K\hat{t}}}\right) \\
	 &= 1 + \tfrac{1}{2}\left(e^{-R} - 1\right)
                        \text{erfc}\left(\frac{\sqrt{R}\hat{x}}{\sqrt{2M\hat{t}}}\right),
  \end{split}
\label{eq4.15}
\end{equation}
where the complementary error function $\text{erfc}(\xi)$ is given in terms of the
error function $\text{erf}(\xi)$ by
\begin{equation}                                
    \text{erfc}(\xi) = 1 - \text{erf}(\xi)
                     = \frac{2}{\sqrt{\pi}}\int_\xi^\infty e^{-\xi'^2} d\xi'.
\label{eq4.16}
\end{equation}
To verify that the boundary conditions in (\ref{eq4.13}) are satisfied, note that
$\text{erfc}(\xi) \to 2$ as $\xi \to -\infty$, and that $\text{erfc}(\xi) \to 0$
as $\xi \to  \infty$. The diffusion equation solution (\ref{eq4.15}) is shown in
the top panel of Fig.~13, where $\varphi(\hat{x},\hat{t})$ is plotted as a function of
$\hat{x}/\sqrt{2M\hat{t}}$ for the three Reynolds' numbers $R=0.3,3,30$.

     Using (\ref{eq4.15}) in (\ref{eq4.14}), we obtain the Burgers' equation
solution
\begin{equation}                                 
   \hat{u}(\hat{x},\hat{t}) = \sqrt{\frac{K}{\pi \hat{t}}}
            \left(\frac{\left(e^{-R} - 1\right)\exp\left(-\frac{\hat{x}^2}{4K\hat{t}}\right)}
                {1+\tfrac{1}{2}\left(e^{-R} - 1\right)
            \text{erfc}\left(\frac{\hat{x}}{\sqrt{4K\hat{t}}}\right)}\right).
\label{eq4.17}
\end{equation}
Translating back to the original variables, the solution (\ref{eq4.17}) can be
written in the form
\begin{equation}                                 
   u(x,t) = U - \sqrt{\frac{M}{2t}}
            \left(\frac{\frac{1}{\sqrt{\pi R}}
	      \left(1-e^{-R}\right)\exp\left(-\frac{R(x-Ut)^2}{2Mt}\right)}
             {1-\tfrac{1}{2}\left(1-e^{-R}\right)
             \text{erfc}\left(\frac{\sqrt{R}(x-Ut)}{\sqrt{2Mt}}\right)}\right).
\label{eq4.18}
\end{equation}
One way to display the solution (\ref{eq4.18}) is to plot $[u(x,t)-U]\sqrt{2t/M}$
as a function of $(x-Ut)/\sqrt{2Mt}$. This is shown in the bottom panel of Fig.~13
for the three different Reynolds' numbers $R=0.3,3,30$. A more physically intuitive
way to display the
solution (\ref{eq4.18}) is to plot $u(x,t)$ for the choices $U=-5$ m s$^{-1}$,
$K=1000$ m$^2$ s$^{-1}$, and $R=30$. This is shown in the top panel of Fig.~14
for $t=2,4,6$ h.
If $u$ is interpreted as the divergent component of the flow in
a slab boundary layer of constant depth $h$, then the implied boundary layer pumping
is given by $w = -h(\partial u/\partial x)$. Profiles of $w(x,t)$ at $t=2,4,6$ h
are shown in the bottom panel of Fig.~14, assuming $h=1000$ m. As discussed in section 2
for the advection equation (see Fig.~6), the shock strength decreases as $t^{-1/2}$
while the width of the subsidence region increases as $t^{1/2}$.
It is worth noting that if $R$ is increased, we will retrieve the asymptotic solutions
shown in Fig.~6 and that if $R$ decreases, the diffusion would increase and smooth
the discontinuity as shown in Fig.~13. The smoothing of the discontinuity in the
$u$ field as $R$ decreases represents a ``shock-like'' feature.

\begin{figure}[!t]             
\centerline{\includegraphics[width=19pc]{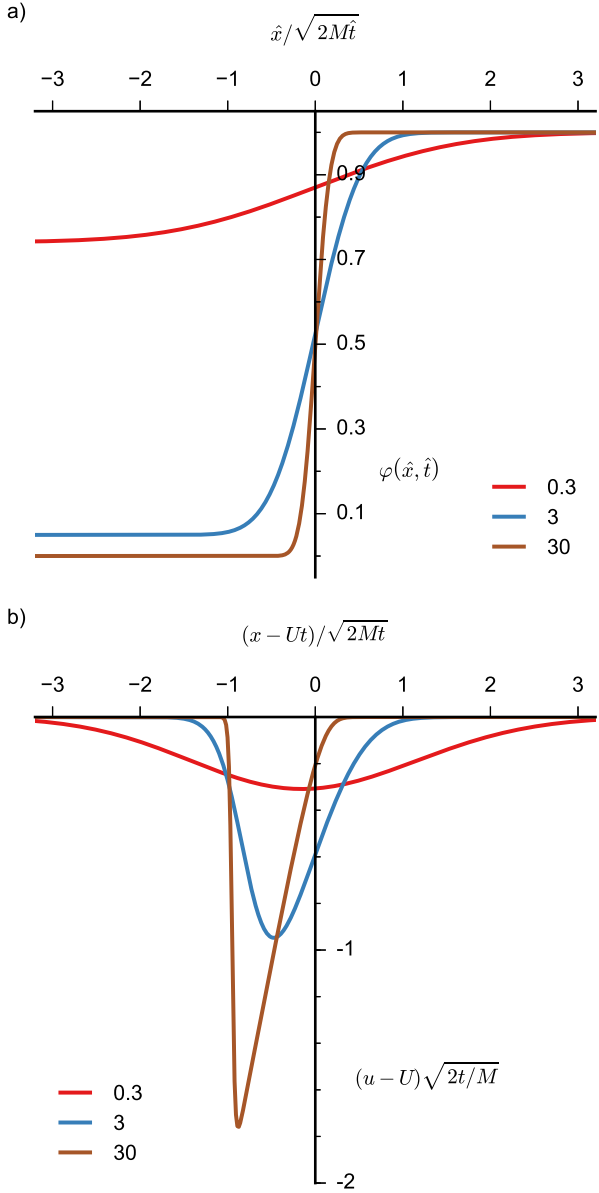}}
\caption{The upper panel shows the diffusion equation solution $\varphi(\hat{x},\hat{t})$,
as given by (\ref{eq4.15}), for the three Reynolds' numbers $R=0.3,3,30$.
The lower panel shows the corresponding Burgers' equation solution, as given by
(\ref{eq4.18}), but plotted with $[u(x,t)-U]\sqrt{2t/M}$ on the ordinate and
$(x-Ut)/\sqrt{2Mt}$ on the abscissa.}
\end{figure}

\begin{figure}[!t]             
\centerline{\includegraphics[width=19pc]{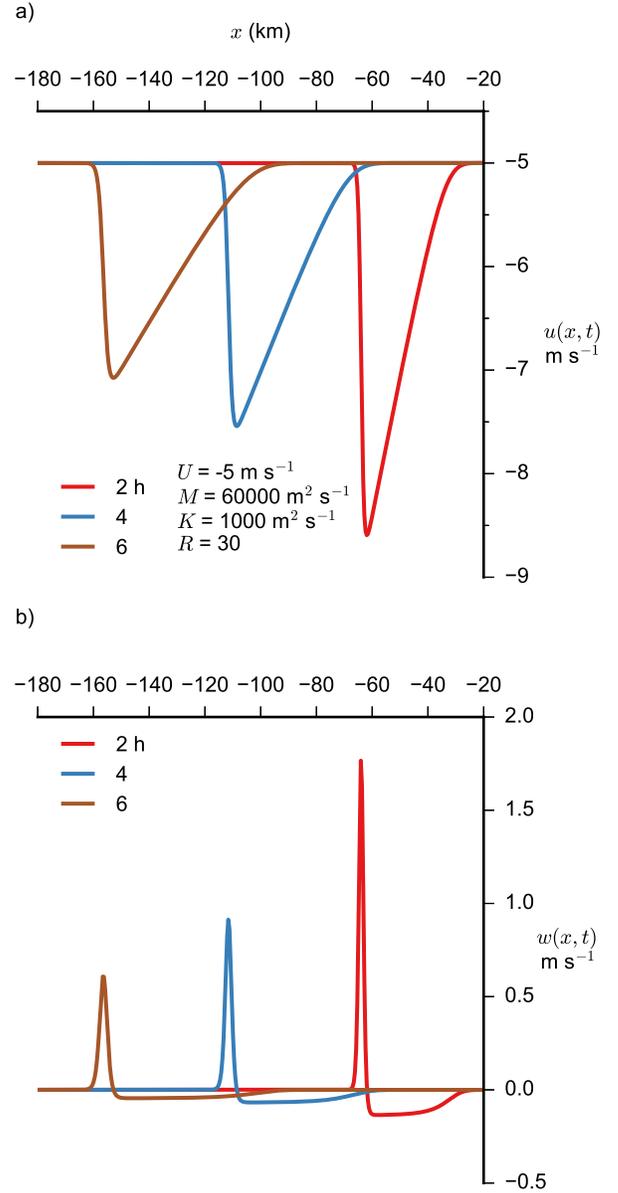}}
\caption{The upper panel shows the Burgers' equation solution (\ref{eq4.18}),
with $u(x,t)$ plotted as a function of $x$ for $t=2,4,6$ h, $U=-5$ m\, s$^{-1}$,
$M=60,000$ m$^2$~s$^{-1}$, and $K=1000$ m$^2$~s$^{-1}$ (so that $R=30$). The lower
panel shows the corresponding $w(x,t)$ field. As this triangular wave
moves to the left, the strength of the shock-like feature on the forward edge decreases as $t^{-1/2}$
while the width of the subsidence region increases as $t^{1/2}$.}
\end{figure}

\section{N-waves, moats, and double eyewalls from Burgers' equation}    

     The diffusion equation solution (\ref{eq4.15}) gives rise to the
triangular wave solution (\ref{eq4.18}). Another interesting diffusion equation
solution gives rise to an N-wave solution. This diffusion equation
solution is
\begin{equation}                                 
     \varphi(\hat{x},\hat{t}) = 1 + \left(\frac{a}{\hat{t}}\right)^{1/2}
                        \exp\left(-\frac{\hat{x}^2}{4K\hat{t}}\right),
\label{eq5.1}
\end{equation}
where the constant $a$ is determined below. Using the diffusion equation
solution (\ref{eq5.1}) in (\ref{eq4.14}), we obtain the Burgers' equation
solution
\begin{equation}                                 
   \hat{u}(\hat{x},\hat{t}) = \frac{\hat{x}}{\hat{t}}
            \left(\frac{\sqrt{a/\hat{t}}\exp\left(-\frac{\hat{x}^2}{4K\hat{t}}\right)}
                     {1+\sqrt{a/\hat{t}}\exp\left(-\frac{\hat{x}^2}{4K\hat{t}}\right)}\right).
\label{eq5.2}
\end{equation}

     The integrated momentum excess in the region $\hat{x}>0$ is defined by
\begin{equation}                                 
  \begin{split}
     M(\hat{t}) &= \int_0^\infty \hat{u}(\hat{x},\hat{t}) \, d\hat{x}
	         = -2K\bigl[\ln\varphi\bigr]_0^\infty  \\
	        &= 2K\ln\left(1+\sqrt{\frac{a}{\hat{t}}}\right),
  \end{split}
\label{eq5.3}
\end{equation}
where the second equality follows from (\ref{eq4.14}) and the third
equality from (\ref{eq5.1}). This is also equal to the integrated momentum deficit
in the region $\hat{x}<0$ as given by
\begin{equation}                                 
  \begin{split}
     M(\hat{t}) &= -\int_{-\infty}^0 \hat{u}(\hat{x},\hat{t}) \, d\hat{x}
	         = 2K\bigl[\ln\varphi\bigr]_{-\infty}^0  \\
	        &= 2K\ln\left(1+\sqrt{\frac{a}{\hat{t}}}\right).
  \end{split}
\label{eq5.4}
\end{equation}
The effective Reynolds' number is defined by
\begin{equation}                                 
     R(\hat{t}) = \frac{M(\hat{t})}{2K} = \ln\left(1+\sqrt{\frac{a}{\hat{t}}}\right),
\label{eq5.5}
\end{equation}
so that the value of $R(\hat{t})$ at $\hat{t}=t_0$ is given by
$R_0 = \ln\left(1+\sqrt{a/t_0}\right)$ and the constant $a$
can be expressed in terms of $R_0$ by $\sqrt{a/t_0} = e^{R_0} - 1$.
Using this last relation, the solution (\ref{eq5.2}) can be written in
the form
\begin{equation}                                 
   \hat{u}(\hat{x},\hat{t}) = \frac{\hat{x}}{\hat{t}}
     \left\{1 + \left(\frac{\hat{t}}{t_0}\right)^{\frac{1}{2}}
     \left(\frac{\exp\left(\frac{\hat{x}^2}{4K\hat{t}}\right)}{e^{R_0}-1}\right)\right\}^{-1}.
\label{eq5.6}
\end{equation}
Translating back to the original variables, we obtain
\begin{equation}                                 
     u(x,t) = U + \frac{x-Ut}{t}
       \left\{1 + \left(\frac{t}{t_0}\right)^{\frac{1}{2}}
       \left(\frac{\exp\left(\frac{(x-Ut)^2}{4Kt}\right)}{e^{R_0} - 1}\right)\right\}^{-1},
\label{eq5.7}
\end{equation}
which is plotted in Fig.~15a for $t=2,4,6$ h.

In order to compare the solution \eqref{eq5.7} to the solution
\eqref{eq3.2} of the nonlinear advection equation in section 3,
consider the case where $R_0 \gg 1$, in which case \eqref{eq5.7} becomes
\begin{equation}                                 
     u(x,t) = U + \frac{x-Ut}{t}
       \left\{1 + \left(\frac{t}{t_0}\right)^{\frac{1}{2}}
       \exp\left(\frac{(x-Ut)^2}{4Kt} - R_0\right)\right\}^{-1}.
\label{eq5.8}
\end{equation}
For this case of $R_0 \gg 1$, in the region $(x-Ut)^2/4Kt < R_0$, the exponential
term in (\ref{eq5.8}) can be neglected so that $u(x,t) \sim x/t$, while in the
region $(x-Ut)^2/4Kt > R_0$, the exponential term is much greater than unity so
that $u(x,t) \sim U$. Thus, the solution is
\begin{equation}                                 
     u(x,t) \sim \begin{cases}
                    U    & \text{if } \qquad\quad -\infty < x < Ut - \sqrt{2Mt} \\
                    x/t  & \text{if }     Ut - \sqrt{2Mt} < x < Ut + \sqrt{2Mt} \\
	            U    & \text{if }     Ut + \sqrt{2Mt} < x < \infty
                 \end{cases}
\label{eq5.9}
\end{equation}
when $R_0$ becomes large.

     As before, if $u$ is interpreted as the divergent component of the flow
in a slab boundary layer of constant depth $h$, then the implied boundary layer
pumping is given by $w=-h(\partial u/\partial x)$, which is plotted in Fig.~15b
for $t=2,4,6$ h. Since the integrated momentum deficit in the region $\hat{x}<0$ is
equal to the integrated momentum excess in the region $\hat{x}>0$, this
example produces shocks of equal strength on the leading and trailing edges
of the widening moat. Examples with shocks of unequal strength are also
possible and examples with the stronger shock on the leading edge more
closely resemble what happens in hurricanes with concentric eyewalls.

\begin{figure}[!t]             
\centerline{\includegraphics[width=19pc]{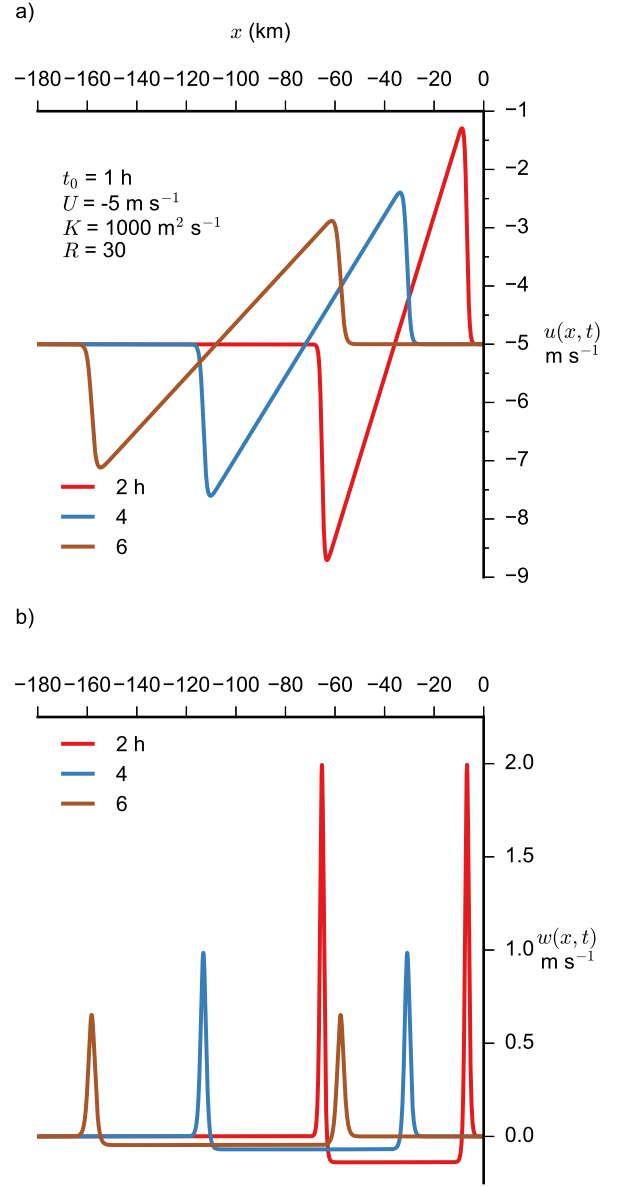}}
\caption{The analytical N-wave solutions for Burgers' equation. The top panel
displays $u(x,t)$ at $t=2,4,6$ h, as computed from (\ref{eq5.7}). These
solutions are for the particular parameters $U=-5$ m s$^{-1}$, $K=1000$ m$^2$~s$^{-1}$,
and $R_0=30$. The bottom panel displays the boundary layer pumping $w(x,t)$, illustrating
the two spikes in vertical motion that surround the widening and subsiding moat region.}
\end{figure}

     The simple Burgers' equation solutions discussed here can serve as the
basis of the following conjecture. When an anomaly forms in the boundary layer
radial inflow, it tends to evolve into either a broadening triangular wave pattern
with concentrated Ekman pumping on the inner edge or a broadening N-wave pattern
with concentrated Ekman pumping on both sides of a moat region with weak subsidence.
In other words, a single eyewall is formed when the $u(\partial u/\partial x)$ term
distorts the boundary layer radial inflow into a triangular wave, while
concentric eyewalls are formed when the inflow is distorted into an N-wave.

\section{Axisymmetric shocks}           

     In the previous two sections, we have studied solutions of the Cartesian
coordinate form of Burgers' equation.  In this section, we shift our attention
to the polar coordinate form
\begin{equation}                                 
        \frac{\partial u}{\partial t}
     + u\frac{\partial u}{\partial r}
     = K\frac{\partial}{\partial r}\left(\frac{\partial(ru)}{r\partial r}\right),
\label{eq6.1}
\end{equation}
which can also be solved analytically using the Cole--Hopf transformation.
In particular, the Cole--Hopf transformation of the nonlinear, advection-diffusion
equation (58) leads to the linear diffusion equation (64). An integral
representation of the solution to (64) is given by (71), where $G(r';r,t)$ is
defined by (73). When this diffusion equation solution $\varphi(r,t)$ is translated
back to the Burgers' equation solution $u(r,t)$, one obtains (74). Since the
Burgers' equation solution (74) is expressed as the ratio of two integrals,
its detailed structure is difficult to see, although the application of asymptotic
methods (not discussed here) can reveal certain aspects of the solution structure.
Because of the mathematical details involved here, some readers may wish to skip
directly to section 7, simply noting that the polar coordinate version (58)
of Burgers' equation can indeed be solved by the Cole--Hopf transformation.

To proceed with the details of the Cole--Hopf transformation,
the first step in the transformation is to write (\ref{eq6.1}) in the form
\begin{equation}                                 
       \frac{\partial u}{\partial t}
     + \frac{\partial}{\partial r}\left(\frac{1}{2}u^2
                                     - K\frac{\partial(ru)}{r\partial r}\right) = 0,
\label{eq6.2}
\end{equation}
and then to define the velocity potential $\chi(r,t)$ such that
\begin{equation}                                 
       u = \frac{\partial\chi}{\partial r},   \qquad
     \frac{1}{2}u^2 - K\frac{\partial(ru)}{r\partial r}
                    = -\frac{\partial\chi}{\partial t}.
\label{eq6.3}
\end{equation}
Combining these last two equations, we obtain
\begin{equation}                                 
                         \frac{\partial\chi}{\partial t}
     +  \frac{1}{2}\left(\frac{\partial\chi}{\partial r}\right)^2
     = K\frac{\partial}{r\partial r}\left(r\frac{\partial\chi}{\partial r}\right).
\label{eq6.4}
\end{equation}
The second step in the Cole--Hopf transformation is to define the new dependent
variable $\varphi(r,t)$ by
\begin{equation}                                 
            \varphi = \exp\left(-\frac{\chi}{2K}\right)
                                 \qquad {\rm or}  \qquad
               \chi = -2K \ln\varphi,
\label{eq6.5}
\end{equation}
from which it follows that
\begin{equation}                                 
         \frac{1}{2}\left(\frac{\partial\chi}{\partial r}\right)^2
     - K \frac{\partial}{r\partial r}\left(r\frac{\partial\chi}{\partial r}\right)
     = \frac{2K^2}{\varphi} \frac{\partial}{r\partial r}
                     \left(r\frac{\partial\varphi}{\partial r}\right).
\label{eq6.6}
\end{equation}
Using (\ref{eq6.6}) in (\ref{eq6.4}), we obtain
\begin{equation}                                 
        \frac{\partial\varphi}{\partial t}
     = K\frac{\partial}{r\partial r}\left(r\frac{\partial\varphi}{\partial r}\right).
\label{eq6.7}
\end{equation}
Thus, the Cole--Hopf procedure has transformed the nonlinear advection-diffusion equation
(\ref{eq6.1}) to the linear diffusion equation (\ref{eq6.7}). If we can solve the diffusion
equation (\ref{eq6.7}) for $\varphi(r,t)$, we can recover the solution of the nonlinear
equation (\ref{eq6.1}) from
\begin{equation}                                 
     u = -\frac{2K}{\varphi} \frac{\partial\varphi}{\partial r}.
\label{eq6.8}
\end{equation}
If $u$ is interpreted as the divergent component of the flow in a slab boundary
layer of constant depth $h$, then the implied boundary layer pumping is given by
\begin{equation}                                 
     w = -h\frac{\partial(ru)}{r\partial r}.
\label{eq6.9}
\end{equation}
In the remainder of this section, we derive solutions of (\ref{eq6.7}) from
which we obtain the corresponding $u$ and $w$ fields.

     Solutions of (\ref{eq6.7}) can be found by a variety of methods,
one of which is the Hankel transform method. The Hankel transform pair
is 
\begin{equation}                                  
   \hat{\varphi}(k,t) = \int_0^\infty \varphi(r,t) J_0(kr) \, r \, dr,
\label{eq6.10}
\end{equation}
\begin{equation}                                  
         \varphi(r,t) = \int_0^\infty \hat{\varphi}(k,t) J_0(kr) \, k \, dk,
\label{eq6.11}
\end{equation}
where $J_0$ is the order zero Bessel function and $k$ is the radial wavenumber.
Multiplying (\ref{eq6.7}) by $rJ_0(kr)$, integrating over all $r$, performing
integration by parts twice  using the Bessel function derivative formulas
$dJ_0(kr)/dr=-kJ_1(kr)$ and $d[rJ_1(kr)]/rdr=kJ_0(kr)$, we can transform the
partial differential equation (\ref{eq6.7}) into the ordinary differential equation
\begin{equation}                                  
       \frac{d\hat{\varphi}}{dt} = -K k^2 \hat{\varphi}.
\label{eq6.12}
\end{equation}
The solution of (\ref{eq6.12}) is
\begin{equation}                                  
  \begin{split}
     \hat{\varphi}(k,t) &= \exp(-Kk^2t) \hat{\varphi}(k,0)        \\
                        &= \exp(-Kk^2t) \int_0^\infty \varphi(r',0) J_0(kr') \, r' \, dr',
\end{split}
\label{eq6.13}
\end{equation}
where the second equality follows from the use of (\ref{eq6.10}) at $t=0$.
Substituting (\ref{eq6.13}) into (\ref{eq6.11}) yields
\begin{equation}                                  
     \varphi(r,t) = \int_0^\infty G(r';r,t) \varphi(r',0)\, r' \, dr'.
\label{eq6.14}
\end{equation}
where
\begin{equation}                                  
          G(r';r,t) = \int_0^\infty  \exp(-Kk^2t) J_0(kr) J_0(kr') \, k \, dk.
\label{eq6.15}
\end{equation}
We next make use of Weber's second exponential integral, which is given on
page 393 of \citet{watson95} and on page 739 of \citet{gradshteyn+ryzhik94}.
This allows us to write (\ref{eq6.15}) as
\begin{equation}                                  
  \begin{split}
     &G(r';r,t) = \frac{1}{2Kt} \exp\left(-\frac{r^2+r'^2}{4Kt}\right) I_0\left(\frac{rr'}{2Kt}\right)  \\
     & \qquad   = \frac{1}{2Kt} \exp\left(-\frac{(r-r')^2}{4Kt}\right)
	                        \exp\left(-\frac{rr'}{2Kt}\right)      I_0\left(\frac{rr'}{2Kt}\right),
  \end{split}
\label{eq6.16}
\end{equation}
where $I_0$ is the order zero modified Bessel function. The second line in (\ref{eq6.16})
is a useful form for $G(r';r,t)$ because $(2\pi x)^{1/2} e^{-x} I_0(x) \to 1$ as $x \to \infty$.

    Equation (\ref{eq6.14}) gives the solution $\varphi(r,t)$ for the diffusion
problem (\ref{eq6.7}). The solution of the original problem (\ref{eq6.1}) is
then found by substituting (\ref{eq6.14}) into (\ref{eq6.8}), which yields
\begin{equation}                                  
     u(r,t) = -2K\left(\frac{\int_0^\infty G_r(r';r,t)\,\varphi(r',0)\, r'\, dr'}
                            {\int_0^\infty G  (r';r,t)\,\varphi(r',0)\, r'\, dr'}\right),
\label{eq6.17}
\end{equation}
where $G_r(r';r,t)$ denotes the partial derivative of $G(r';r,t)$ with respect to $r$.
Note that in the relation (\ref{eq6.16}) for $G(r';r,t)$, and thus in (\ref{eq6.14}) for
$\varphi(r,t)$, the constant $K$ always appears coupled to $t$, i.e., only as the product $Kt$.
This is not a property of the $u(r,t)$ solution (\ref{eq6.17}).
In fact, the $\varphi(r,t)$ field diffuses while the $u(r,t)$ field shocks.

     As a simple example, choose the initial $\varphi$ field to be
\begin{equation}                                  
  \begin{split}
     \varphi(r,0) = &\left[1 + (n_1-1)\left(\frac{r}{a_1}\right)^{n_1}\right]^{\frac{a_1 U_1}{2K(n_1-1)}} \\
              \cdot &\left[1 + (n_2-1)\left(\frac{r}{a_2}\right)^{n_2}\right]^{\frac{a_2 U_2}{2K(n_2-1)}},
  \end{split}
\label{eq6.18}
\end{equation}
where $a_1,a_2,n_1,n_2,U_1,U_2$ are constants. The exponents in (\ref{eq6.18}) define two
Reynolds' numbers as $R_1=a_1U_1/(2K)$ and $R_2=a_2U_2/(2K)$. For example, if $a_1=20$ km,
$U_1=10$ m s$^{-1}$, $a_2=40$ km, $U_2=20$ m s$^{-1}$, and $K=1000$ m$^2$ s$^{-1}$,
we have $R_1=100$ and $R_2=400$.
Using (\ref{eq6.5}), the corresponding initial $\chi$ field is
\begin{equation}                                  
  \begin{split}
     \chi(r,0) = &-\frac{a_1 U_1}{n_1-1} \ln\left[1 + (n_1-1)\left(\frac{r}{a_1}\right)^{n_1}\right] \\
                 &-\frac{a_2 U_2}{n_2-1} \ln\left[1 + (n_2-1)\left(\frac{r}{a_2}\right)^{n_2}\right],
  \end{split}
\label{eq6.19}
\end{equation}
and, from the first entry in (\ref{eq6.3}), the corresponding initial $u$ field is
\begin{equation}                                  
  \begin{split}
     u(r,0) = &-U_1\left(\frac{n_1(r/a_1)^{n_1-1}}{1 + (n_1-1)(r/a_1)^{n_1}}\right) \\
	      &-U_2\left(\frac{n_2(r/a_2)^{n_2-1}}{1 + (n_2-1)(r/a_2)^{n_2}}\right).
  \end{split}
\label{eq6.20}
\end{equation}
For reasonable choices such as $n_1=4$ and $n_2=8$, this example illustrates a simple boundary layer
mechanism for the merging of tropical cyclone convective rings into a single eyewall structure.
In other words, the solution (\ref{eq6.17}) can describe the merger of two shocks that
propagate inward. As the outer shock overtakes the inner one,
the details of the evolving structure are lost and a very simple final shock-like
structure is obtained. A thorough examination of such solutions is left for future study.

     We conclude Part I by asking: ``How do hurricane eyewalls originate?"
The results of sections 2 and 4 suggest the possibility that a single eyewall
is a phenomenon instigated by the tendency of the boundary layer radial
inflow to form a single shock on the inward edge of a region of enhanced
radial inflow. Similarly, the results of sections 3 and 5 suggest the possibility
that a double eyewall is a phenomenon instigated by the tendency of the
boundary layer radial inflow to form a double shock (or N-wave) on the
inward and outward edges of a region that has both enhanced and reduced
radial inflow. In either case, the formation of a boundary layer shock may
be one of the most important events in the life cycle of a hurricane, for
it imposes on the storm a classic eye/eyewall structure. In comparing our
solutions to observed aspects of the tropical cyclone boundary layer, we note
that these solutions lack the pressure gradient force and dissipative effects.
The result is that we capture only some of the evolution seen in nature.

\bigskip

\centerline{\bf II. Line-Symmetric Slab Ekman Layer Model}

\section{Analytical solutions for $y$-independent shocks} 

     We now return to the discussion of the slab boundary layer model
(\ref{eq1.1}). In the absence of horizontal diffusion, the line-symmetric
version of (\ref{eq1.1}) can be solved analytically using the method of
characteristics. Thus, consider the line-symmetric slab boundary layer equations
\begin{equation}                                  
  \begin{split}
     \frac{\partial u}{\partial t} + u\frac{\partial u}{\partial x}
            - fv + \frac{\cD U}{h}u &= -\frac{1}{\rho}\frac{\partial p}{\partial x},  \\
     \frac{\partial v}{\partial t} + u\frac{\partial v}{\partial x}
            + fu + \frac{\cD U}{h}v &= 0,
  \end{split}
\label{eq7.1}
\end{equation}
where $U=(u^2+v^2)^{1/2}$ is the wind speed, and where the Coriolis parameter $f$, the
boundary layer depth $h$, and the drag coefficient $\cD$ are assumed to have the
values $f=5\times 10^{-5}$ s$^{-1}$, $h=1000$ m, and $\cD=2\times 10^{-3}$. The forcing
term $-(1/\rho)(\partial p/\partial x)$, which is also assumed to be a constant, can
be interpreted in terms of a specified geostrophic wind $v_g$, since
$fv_g=(1/\rho)(\partial p/\partial x)$.  Our goal is to solve the system (\ref{eq7.1})
for $u(x,t)$ and $v(x,t)$ on an infinite domain subject to the initial conditions
\begin{equation}                                  
  \begin{split}
        u(x,0) = u_0(x)  \quad \text{and} \quad  v(x,0) = v_0(x),
  \end{split}
\label{eq7.2}
\end{equation}
where $u_0(x)$ and $v_0(x)$ are specified functions. Obviously, the line-symmetric
boundary layer dynamics (\ref{eq7.1}) lacks important curvature effects and misses
important spatial variations to the forcing that are
present in the axisymmetric dynamics (\ref{eq1.1}). Thus, (\ref{eq7.1}) should
 be regarded as a qualitative model of the hurricane boundary layer. Its
attraction is the ease with which analytical solutions can be obtained
for a coupled pair of equations that increases our understanding of shocks in the
full slab model \eqref{eq1.1}.

     The quasi-linear system (\ref{eq7.1}) is hyperbolic and
can be written in characteristic form, i.e., it can be written as a system of
ordinary differential equations. In order to make the characteristic form of
the $u$ and $v$ equations homogeneous, it is convenient to introduce the constant
Ekman flow components $u_{_E}$ and $v_{_E}$, which are determined from the nonlinear
algebraic system
\begin{equation}                                  
  \begin{split}
      - fv_{_E} + \frac{\cD U_{_E}}{h}u_{_E} &= -fv_g,   \\
        fu_{_E} + \frac{\cD U_{_E}}{h}v_{_E} &= 0,
  \end{split}
\label{eq7.3}
\end{equation}
where $U_{_E}=(u_{_E}^2+v_{_E}^2)^{1/2}$. The ``solutions" of (\ref{eq7.3}) are
\begin{equation}                                  
  \begin{split}
         u_{_E} &= -\left(\frac{f(\cD U_{_E}/h)}{f^2 + (\cD U_{_E}/h)^2}\right)v_g, \\
         v_{_E} &=  \left(\frac{f^2            }{f^2 + (\cD U_{_E}/h)^2}\right)v_g.
  \end{split}
\label{eq7.4}
\end{equation}
These two relations are implicit because the wind speed $U_{_E}$ depends on
the velocity components $u_{_E}$ and $v_{_E}$.
However, we can find an explicit solution for $\cD U_{_E}/h$ by
squaring each equation in (\ref{eq7.4}) and adding the results to obtain
\begin{equation}                                 
         k^2 = \left(\frac{f^2}{f^2 + k^2}\right) k_g^2,
\label{eq7.5}
\end{equation}
where $k=\cD U_{_E}/h$ and $k_g=\cD v_g/h$. Equation (\ref{eq7.5}) can be solved as a
quadratic for $k^2$, yielding\footnote{Note that $k/f$ can be interpreted as the ``slab
Ekman number," i.e., as the ratio of the magnitudes of the drag force and the Coriolis
force. Similarly, $k_g/f$ can be interpreted as the ``forced Ekman number."}
\begin{equation}                                  
    \frac{k}{f} = \left\{\left[\frac{1}{4} + \left(\frac{k_g}{f}\right)^2\right]^{1/2}
                             - \frac{1}{2}\right\}^{1/2}.
\label{eq7.6}
\end{equation}
For the five values of $v_g$ listed in the first column of Table 2, the second column
lists the corresponding values of $k_g/f$, the third column lists the corresponding values
of $k/f$ determined from (\ref{eq7.6}), while the fourth and fifth columns list
the corresponding values of $u_{_E}$ and $v_{_E}$ determined from (\ref{eq7.4}).

\begin{table}[b!t]                        
\centering
{\begin{tabular}{ccccc}
\hline\hline 
     $v_g$   & $k_g/f$ & $k/f$  & $u_{_E}$   & $v_{_E}$    \\
 (m s$^{-1}$)&         &        &(m s$^{-1}$)&(m s$^{-1}$) \\
\hline 
       10    &  0.4    & 0.3746 &  $ -3.29$  &     8.77    \\
       20    &  0.8    & 0.6659 &  $ -9.23$  &    13.86    \\
       30    &  1.2    & 0.8944 &  $-14.91$  &    16.67    \\
       40    &  1.6    & 1.0846 &  $-19.93$  &    18.38    \\
       50    &  2.0    & 1.2496 &  $-24.39$  &    19.52    \\
\hline 
\end{tabular}}
\caption{The geostrophic wind $v_g$, the corresponding ``forced Ekman number"
$k_g/f$, the ``slab Ekman number" (or dimensionless damping rate) $k/f$,
and the steady-state Ekman layer components $u_{_E}$ and $v_{_E}$ for five selected
cases.}
\end{table}

\begin{figure}[b!t]                          
\centerline{\includegraphics[width=19pc]{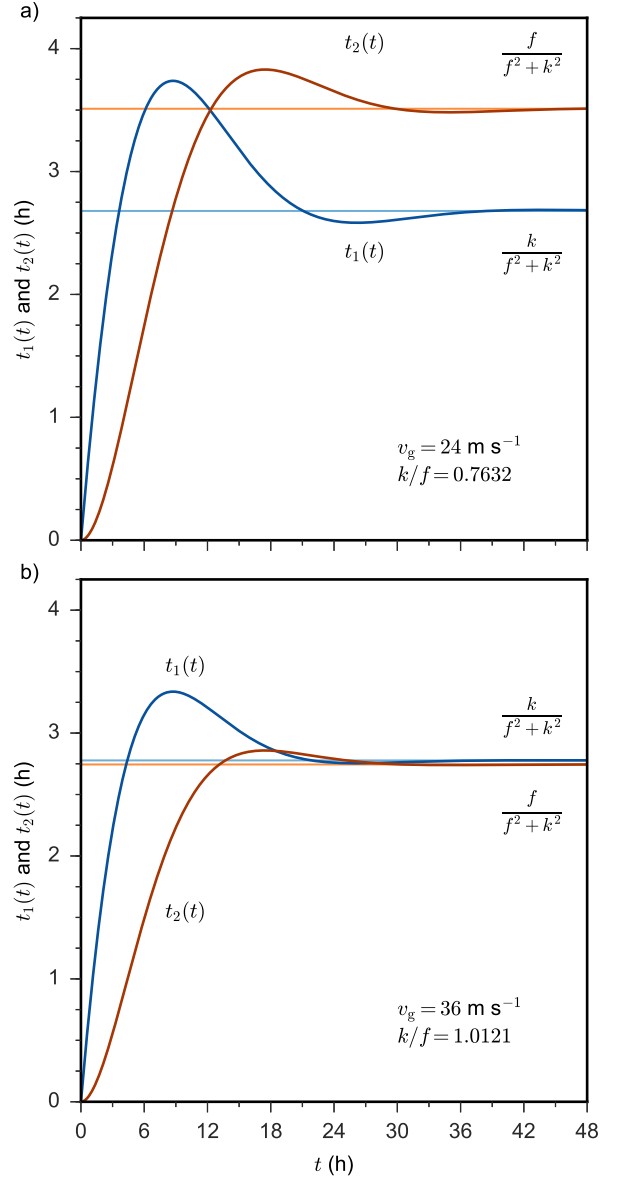}}
\caption{The functions $t_1(t)$ (blue curves) and $t_2(t)$ (red curves)
for a) $k/f = 0.7632$ ($v_g=24$ m s$^{-1}$) and b) $k/f = 1.0121$ ($v_g=36$ m s$^{-1}$).
Note that $t_1(t)\rightarrow k/(f^2+k^2)$
(horizontal blue lines) and $t_2(t)\rightarrow f/(f^2+k^2)$ (horizontal red lines)
as $t \rightarrow\infty$.  The maximum value of $t_1(t)$ occurs at $t=\pi/(2f)\approx 8.7$ h,
while the maximum value of $t_2(t)$ occurs at $t=\pi/f\approx 17.5$ h.}
\end{figure}

    We now approximate the $\cD U$ factors in (\ref{eq7.1}) by $\cD U_{_E}$. Then,
combining this approximate form of (\ref{eq7.1}) with (\ref{eq7.3}), we obtain
the characteristic form
\begin{equation}                                  
     \left.\begin{matrix}
                \displaystyle{\frac{d(u-u_{_E})}{dt} - f(v-v_{_E}) + k(u-u_{_E}) = 0} \\[1.5ex]
                \displaystyle{\frac{d(v-v_{_E})}{dt} + f(u-u_{_E}) + k(v-v_{_E}) = 0}
	   \end{matrix}\right\} \,\, \text{on} \,\, \frac{dx}{dt} = u,
\label{eq7.7}
\end{equation}
where $(d/dt)=(\partial/\partial t)+u(\partial/\partial x)$ can be interpreted as
the derivative along a characteristic. In the special case $f=0$, the two momentum
equations in (\ref{eq7.7}) decouple, and the first reduces to the nonlinear advection
equation with damping, which was discussed in section 3b.
Thus, we anticipate the possible appearance of shocks in the solutions of the coupled
equations (\ref{eq7.7}). As can be checked by direct substitution, the solutions
of the coupled $u$ and $v$ equations in (\ref{eq7.7}) are
\begin{equation}                                  
  \begin{split}
       u(x,t) = u_{_E} &+ [u_0(\hat{x})-u_{_E}]e^{-kt}\cos(ft)  \\
                       &+ [v_0(\hat{x})-v_{_E}]e^{-kt}\sin(ft),
  \end{split}
\label{eq7.8}
\end{equation}
\begin{equation}                                  
  \begin{split}
       v(x,t) = v_{_E} &- [u_0(\hat{x})-u_{_E}]e^{-kt}\sin(ft)  \\
                       &+ [v_0(\hat{x})-v_{_E}]e^{-kt}\cos(ft),
  \end{split}
\label{eq7.9}
\end{equation}
where $\hat{x}$ is the initial position of the characteristic.
According to (\ref{eq7.8}) and (\ref{eq7.9}), if $u_0(x) \ne u_{_E}$ and/or
$v_0(x) \ne v_{_E}$, there will be damped inertial oscillations along each
characteristic, leading to eventual steady-state Ekman balance.
To find the shapes of the characteristics, we now substitute the solution for
$u(x,t)$ into the right-hand side of $(dx/dt)=u$, and then integrate from zero
to $t$ along a characteristic, thereby obtaining
\begin{equation}                                  
       x = \hat{x} + u_{_E}t + [u_0(\hat{x})-u_{_E}]t_1(t) + [v_0(\hat{x})-v_{_E}]t_2(t),
\label{eq7.10}
\end{equation}
where the $t_1(t)$ and $t_2(t)$ functions are defined by
\begin{equation}                                  
  \begin{split}
        t_1(t) &= \frac{k - e^{-kt}[k\cos(ft) - f\sin(ft)]}{f^2 + k^2},   \\
        t_2(t) &= \frac{f - e^{-kt}[k\sin(ft) + f\cos(ft)]}{f^2 + k^2}.
  \end{split}
\label{eq7.11}
\end{equation}
To verify that (\ref{eq7.10}) and (\ref{eq7.11}) constitute a solution of $(dx/dt)=u$,
take $d/dt$ of (\ref{eq7.10}) and make use of $(dt_1/dt)=e^{-kt}\cos(ft)$ and
$(dt_2/dt)=e^{-kt}\sin(ft)$.
Plots of $t_1(t)$ and $t_2(t)$ for the cases $k/f=0.7632$ ($v_g=24$  m s$^{-1}$)
and $k/f=1.0121$ ($v_g=36$  m s$^{-1}$) are shown
in the two panels of Fig.~16. Since each characteristic can be
considered to be uniquely labeled by its value of $\hat{x}$, equations (\ref{eq7.8})
and (\ref{eq7.9}) give the variation of $u$ and $v$ along the characteristic, while
(\ref{eq7.10}) gives the shape of the characteristic. Thus, (\ref{eq7.8})--(\ref{eq7.11})
constitute the solution of the original problem (\ref{eq7.1}) and (\ref{eq7.2}), with
the understanding that the $\cD U$ factors in (\ref{eq7.1}) have been approximated
by $\cD U_{_E}$, and the solution does not extend past shock formation
time.\footnote{In analogy with the procedure used in section 3, iterative calculations
can be avoided in dealing with the implicit nature of the solutions
(\ref{eq7.8})--(\ref{eq7.11}) by producing plots of these solutions as follows.
Choose a time $t$ and then calculate the corresponding $t_1(t)$ and $t_2(t)$
from (\ref{eq7.11}). Choose a set of equally spaced values of $\hat{x}$ and then
use (\ref{eq7.10}) to calculate the corresponding set of unequally spaced values
of $x$. Then use (\ref{eq7.8}) and (\ref{eq7.9}) to calculate $u(x,t)$ and $v(x,t)$
at the unequally spaced $x$-points. Finally, plot $u(x,t)$ and $v(x,t)$ as functions
of $x$ at the chosen time $t$ using a plotting routine that can handle unequally spaced
data points.}

    To understand when the divergence $\delta=(\partial u/\partial x)$ and the
vorticity $\zeta=(\partial v/\partial x)$ become infinite, we first note that
$(\partial/\partial x)$ of (\ref{eq7.10}) yields
\begin{equation}                                  
     \frac{\partial\hat{x}}{\partial x} = \frac{1}{1 + t_1(t)\delta_0(\hat{x}) + t_2(t)\zeta_0(\hat{x})},
\label{eq7.12}
\end{equation}
so that $(\partial/\partial x)$ of (\ref{eq7.8}) and (\ref{eq7.9}) yield
\begin{equation}                                  
  \begin{split}
     \delta(x,t) &= \frac{e^{-kt}\left[ \delta_0(\hat{x})\cos(ft) + \zeta_0(\hat{x})\sin(ft)\right]}
	                 {1 + t_1(t)\delta_0(\hat{x}) + t_2(t)\zeta_0(\hat{x})},     \\
     \zeta(x,t)  &= \frac{e^{-kt}\left[-\delta_0(\hat{x})\sin(ft) + \zeta_0(\hat{x})\cos(ft)\right]}
                         {1 + t_1(t)\delta_0(\hat{x}) + t_2(t)\zeta_0(\hat{x})}.
  \end{split}
\label{eq7.13}
\end{equation}
From the analytical solutions (\ref{eq7.13}), we can easily obtain
\begin{equation}                                    
        \left(\frac{\delta^2(x,t) + \zeta^2(x,t)}
	           {\delta_0^2(\hat{x}) + \zeta_0^2(\hat{x})}\right)^{1/2}
     = \frac{e^{-kt}}
            {1 + t_1(t)\delta_0(\hat{x}) + t_2(t)\zeta_0(\hat{x})}.
\label{eq7.14}
\end{equation}
To compute the time of shock formation, we note that, from the denominators on the
right-hand sides of (\ref{eq7.13}), the divergence and the vorticity can become
infinite if
\begin{equation}                                  
          t_1(t)\delta_0(\hat{x}) + t_2(t)\zeta_0(\hat{x}) = -1
\label{eq7.15}
\end{equation}
along one or more of the characteristics.  Note from Fig.~16 that the values of $t_1$
and $t_2$ may never get large enough to satisfy (\ref{eq7.15}), in which case a shock
will not form. In section 9 we consider shock formation for initial conditions with
$\zeta_0(x)=0$ and $\delta_0(x)\ne 0$, while in section 10 we consider initial conditions
with $\delta_0(x)=0$ and $\zeta_0(x)\ne 0$. However, before discussing these particular
initial conditions, we provide in section 8 an alternative derivation of the solutions
(\ref{eq7.13}).

\section{Alternative derivation of the $\delta$ and $\zeta$ solutions}   

    Since it is the divergence and vorticity that can become infinite when a shock occurs, rather than
the velocity components $u$ and $v$, it is of interest to recall the governing equations
for $\delta$ and $\zeta$. These equations, derived from (\ref{eq7.7}), can be written as
\begin{equation}                                    
     \frac{d\delta}{dt} + \delta^2 - f\zeta + k\delta = 0,
\label{eq8.1}
\end{equation}
\begin{equation}                                    
     \frac{d\zeta}{dt} + (f + \zeta)\delta + k\zeta = 0,
\label{eq8.2}
\end{equation}
where we have assumed $(\partial u_{_E}/\partial x)=0$ and
$(\partial v_{_E}/\partial x)=0$. Note that the $\delta^2$ term
in (\ref{eq8.1}) originates from the $u(\partial u/\partial x)$ term in the
$x$-momentum equation and that the $\zeta\delta$ term in (\ref{eq8.2}) originates
from the $u(\partial v/\partial x)$ term in the $y$-momentum equation. Thus, we
expect that the $\delta^2$ and $\zeta\delta$ terms play a crucial role in the
development of any singularities in divergence and vorticity.

     Taking the sum of $\delta$ times (\ref{eq8.1}) and $\zeta$ times (\ref{eq8.2}),
we obtain
\begin{equation}                                    
         \frac{d}{dt}\left(\delta^2 + \zeta^2\right)^{1/2}
      = -(\delta + k)\left(\delta^2 + \zeta^2\right)^{1/2},
\label{eq8.3}
\end{equation}
so that $(\delta^2+\zeta^2)^{1/2}$ decays along a characteristic when $-\delta < k$
and grows along a characteristic when $-\delta > k$, i.e., growth occurs when
the magnitude of convergence exceeds the critical value $k$.
If, at any time $t$, the divergence and vorticity satisfy $(\delta^2+\zeta^2)^{1/2} < k$,
then it follows that $\delta + k > 0$ and, according to (\ref{eq8.3}),
$(\delta^2+\zeta^2)^{1/2}$ will further decrease. Thus, a necessary condition
for shock formation is $(\delta_0^2+\zeta_0^2)^{1/2} > k$.

\begin{figure}[!th]                          
\centerline{\includegraphics[width=19pc]{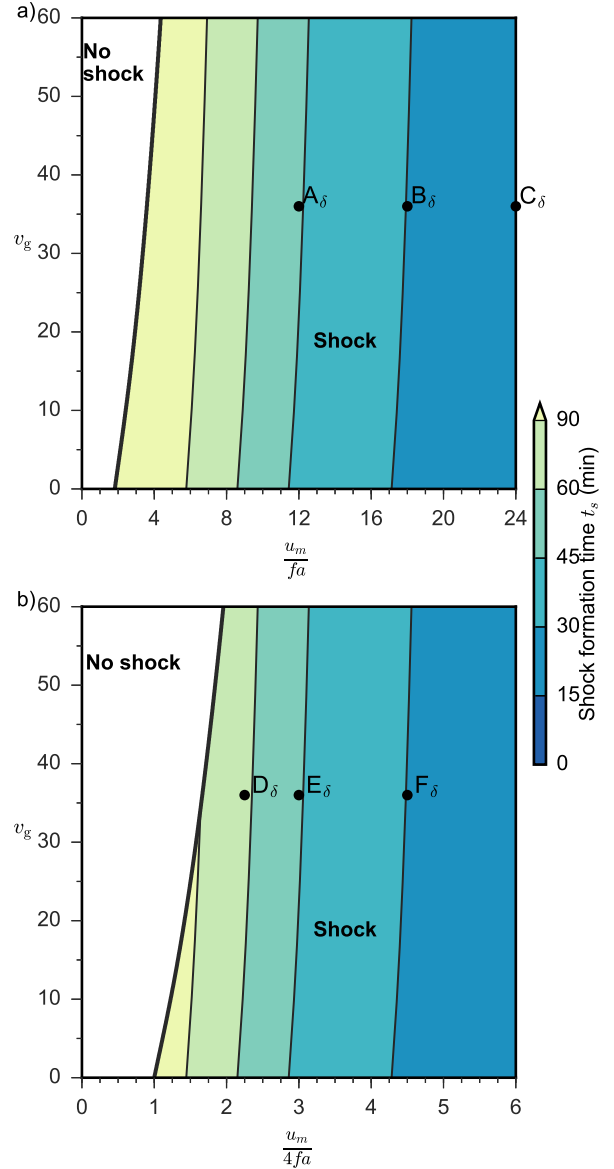}}
\caption{Isolines of the shock formation time $t_s$ (color shading) and the shock
condition (thick curve) for a) the initial divergence cases resulting in triangular
waves, as described by equations (\ref{eq9.8}) and (\ref{eq9.10}) and b) the
initial divergence cases resulting in N-waves, as described by
equations (\ref{eq9.18}) and (\ref{eq9.20}). The abscissa in each panel is
a dimensionless measure of the maximum initial convergence. The points
$A_\delta,B_\delta,C_\delta$ in the upper panel correspond to the three columns
in Fig.~18, while the points $D_\delta,E_\delta,F_\delta$ in the lower panel
correspond to the three columns in Fig.~19.}
\end{figure}

    We now consider the derivation of the solutions for $\delta$ and $\zeta$
directly from (\ref{eq8.1}) and (\ref{eq8.2}). As before, let
$\hat{x}$ be the label of a given characteristic, with a convenient choice
for this label being the initial position of the characteristic. Since the label
is invariant along the characteristic, we have $(d\hat{x}/dt)=0$. Then, taking
$(\partial/\partial x)$ of this last relation we obtain
\begin{equation}                                    
            \frac{d}{dt}\left(\frac{\partial\hat{x}}{\partial x}\right)
         = -\delta \left(\frac{\partial\hat{x}}{\partial x}\right),
\label{eq8.4}
\end{equation}
which is an equation relating the spacing of the characteristics
$(\partial\hat{x}/\partial x)^{-1}$ to the divergence $\delta$. Now search for
solutions of (\ref{eq8.1}) and (\ref{eq8.2}) having the form
\begin{equation}                                    
  \begin{split}
       \delta(x,t) &= \tilde{\delta}(x,t)\left(\frac{\partial\hat{x}}{\partial x}\right), \\
       \zeta (x,t) &= \tilde{\zeta }(x,t)\left(\frac{\partial\hat{x}}{\partial x}\right).
  \end{split}
\label{eq8.5}
\end{equation}
When characteristics come together in the $(x,t)$-plane, the dimensionless spacing
$(\partial\hat{x}/\partial x)^{-1}$ goes to zero and the above factor
$(\partial\hat{x}/\partial x)$ goes to infinity. In this way the solutions $\delta$
and $\zeta$ can become singular while $\tilde{\delta}$ and $\tilde{\zeta}$ remain
well-behaved. Substitution of (\ref{eq8.5}) into the divergence equation (\ref{eq8.1})
yields
\begin{equation}                                    
  \begin{split}
    0&= \frac{d\delta}{dt} + \delta^2 - f\zeta + k\delta      \\
     &= \left(\frac{\partial\hat{x}}{\partial x}\right)
        \left(\frac{d\tilde{\delta}}{dt} - f\tilde{\zeta} + k\tilde{\delta}\right)
      + \tilde{\delta} \frac{d}{dt}\left(\frac{\partial\hat{x}}{\partial x}\right)
      + \tilde{\delta}^2\left(\frac{\partial\hat{x}}{\partial x}\right)^2     \\
     &= \left(\frac{\partial\hat{x}}{\partial x}\right)
        \left(\frac{d\tilde{\delta}}{dt} - f\tilde{\zeta} + k\tilde{\delta}\right),
  \end{split}
\label{eq8.6}
\end{equation}
where the last line follows from the fact that (\ref{eq8.4}) can be used to verify
cancellation of the last two terms in the second line.
Similarly, substitution of (\ref{eq8.5}) into the vorticity equation (\ref{eq8.2})
yields
\begin{equation}                                    
  \begin{split}
    0&= \frac{d\zeta}{dt} + \delta\zeta + f\delta + k\zeta      \\
     &= \left(\frac{\partial\hat{x}}{\partial x}\right)
        \left(\frac{d\tilde{\zeta}}{dt} + f\tilde{\delta} + k\tilde{\zeta}\right)
      + \tilde{\zeta} \frac{d}{dt}\left(\frac{\partial\hat{x}}{\partial x}\right)
      + \tilde{\delta}\tilde{\zeta}\left(\frac{\partial\hat{x}}{\partial x}\right)^2     \\
     &= \left(\frac{\partial\hat{x}}{\partial x}\right)
        \left(\frac{d\tilde{\zeta}}{dt} + f\tilde{\delta} + k\tilde{\zeta}\right),
  \end{split}
\label{eq8.7}
\end{equation}
where, as before, the last line follows from the fact that (\ref{eq8.4}) can be used
to verify cancellation of the last two terms in the second line.
Thus, while $\delta$ and $\zeta$ satisfy the nonlinear equations (\ref{eq8.1})
and (\ref{eq8.2}), the new variables $\tilde{\delta}$ and $\tilde{\zeta}$
satisfy the linear equations
\begin{equation}                                    
  \begin{split}
       \frac{d\tilde{\delta}}{dt} &- f\tilde{\zeta}  + k\tilde{\delta} = 0,    \\
       \frac{d\tilde{\zeta} }{dt} &+ f\tilde{\delta} + k\tilde{\zeta}  = 0.
  \end{split}
\label{eq8.8}
\end{equation}
The solutions of the coupled equations (\ref{eq8.8}) are
\begin{equation}                                    
  \begin{split}
     \tilde{\delta}(x,t) &= e^{-kt}\left[\,\delta_0(\hat{x})\cos(ft)
                                         + \zeta_0 (\hat{x})\sin(ft)\right],  \\
     \tilde{\zeta} (x,t) &= e^{-kt}\left[- \delta_0(\hat{x})\sin(ft)
                                         + \zeta_0 (\hat{x})\cos(ft)\right].
  \end{split}
\label{eq8.9}
\end{equation}
Combining (\ref{eq8.4}), (\ref{eq8.5}), and the first line of (\ref{eq8.9}), it
can be shown that
\begin{equation}                                    
     \frac{d}{dt}\left[\left(\frac{\partial\hat{x}}{\partial x}\right)^{-1}\right]
         = e^{-kt}\left[\,\delta_0(\hat{x})\cos(ft)
                        + \zeta_0 (\hat{x})\sin(ft)\right].
\label{eq8.10}
\end{equation}
Integrating (\ref{eq8.10}), noting that $(\partial\hat{x}/\partial x)=1$ at $t=0$,
we find that the spacing of the characteristics is given by the inverse of
(\ref{eq7.12}), i.e.,
\begin{equation}                                    
     \left(\frac{\partial\hat{x}}{\partial x}\right)^{-1}
     = 1 + t_1(t)\delta_0(\hat{x}) + t_2(t)\zeta_0(\hat{x}).
\label{eq8.11}
\end{equation}
When (\ref{eq8.9}) and (\ref{eq8.11}) are substituted into (\ref{eq8.5}), we
recover the previously derived solutions (\ref{eq7.13}) for $\delta$ and $\zeta$
and gain insight into the role of the intersection of characteristics in the
formation of singularities in $\delta$ and $\zeta$.

\section{Examples with initial divergence only}  

     In this section, we consider examples for which there is initial
divergence, but no initial vorticity. The initial condition used in
section 9a leads to the formation of a triangular wave, or single
eyewall structure, while the initial condition used in section 9b
leads to the formation of an N-wave, or double eyewall structure.
In section 10, we consider examples for which there is initial vorticity,
but no initial divergence.

\subsection{Formation of a triangular wave}

\begin{figure*}[!t]                   
\centerline{\includegraphics[width=39pc]{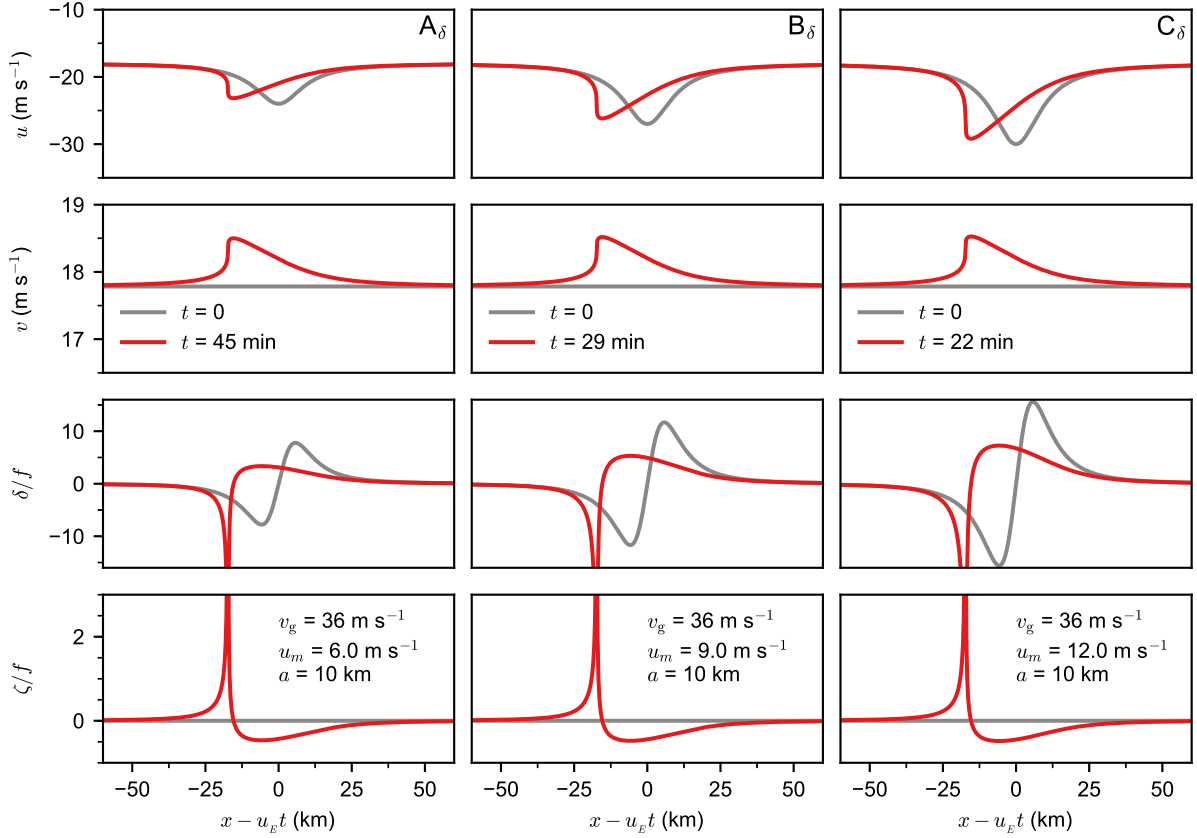}}
\caption{The three columns show three examples with initial divergence only,
as determined by the analytical solutions (\ref{eq9.3})--(\ref{eq9.5}). All three
lead to the formation of triangular waves in $u$ and $v$. The
spatial distributions at $t=0$ are shown by the gray curves, while the distributions
at shock formation time are shown by the red curves. All three cases have
$v_g=36$ m s$^{-1}$, $a=10$ km, and the same initial $v_0(x)=v_{_E}=17.8$ m~s$^{-1}$,
so that the initial vorticity is zero. These three examples correspond to the three
points labeled $A_\delta,B_\delta,C_\delta$ in the top panel of Fig.~17.}
\end{figure*}

     In the first example, consider the initial conditions
\begin{equation}                                  
       u_0(x) = u_{_E} - u_m \left(\frac{1}{1 + (x/a)^2}\right)
                \quad \text{and} \quad v_0(x) = v_{_E},
\label{eq9.1}
\end{equation}
where the constants $a$ and $u_m$ specify the horizontal extent and strength
of this initial symmetric divergent flow anomaly. The initial divergence and
vorticity associated with (\ref{eq9.1}) are
\begin{equation}                                  
     \delta_0(x) = \frac{2u_m}{a} \left(\frac{x/a}{[1 + (x/a)^2]^2}\right)
          \quad \text{and} \quad  \zeta_0(x)  = 0.
\label{eq9.2}
\end{equation}
We assume $u_m>0$ so that initially convergence appears to the left
of the origin and divergence to the right. With these initial conditions,
the solutions (\ref{eq7.8}) and (\ref{eq7.9}) simplify to
\begin{equation}                                  
  \begin{split}
       u(x,t) &= u_{_E} - u_m\left(\frac{1}{1 + (\hat{x}/a)^2}\right)e^{-kt}\cos(ft), \\
       v(x,t) &= v_{_E} + u_m\left(\frac{1}{1 + (\hat{x}/a)^2}\right)e^{-kt}\sin(ft),
  \end{split}
\label{eq9.3}
\end{equation}
while the characteristic equation (\ref{eq7.10}) simplifies to
\begin{equation}                                  
       x = \hat{x} + u_{_E}t - u_m\left(\frac{1}{1 + (\hat{x}/a)^2}\right)t_1(t).
\label{eq9.4}
\end{equation}
The solutions (\ref{eq7.13}) for the divergence and vorticity become
\begin{equation}                                  
       \delta(x,t) = \frac{\delta_0(\hat{x})e^{-kt}\cos(ft)}{1 + t_1(t)\delta_0(\hat{x})},   \quad
       \zeta (x,t) =-\frac{\delta_0(\hat{x})e^{-kt}\sin(ft)}{1 + t_1(t)\delta_0(\hat{x})},
\label{eq9.5}
\end{equation}
so that
\begin{equation}                                    
        \left(\frac{\delta^2(x,t) + \zeta^2(x,t)}{\delta_0^2(\hat{x})}\right)^{1/2}
     = \frac{e^{-kt}}{1 + t_1(t)\delta_0(\hat{x})}.
\label{eq9.6}
\end{equation}
From (\ref{eq9.5}) or (\ref{eq9.6}), shock formation occurs along the characteristic
$\hat{x}$ when $t_1(t)\delta_0(\hat{x}) = -1$. This occurs first along the characteristic
with the minimum value of $\delta_0(\hat{x})$. For this example, the minimum value of
$\delta_0(\hat{x})$ occurs at $\hat{x}=-a/\sqrt{3}\equiv\hat{x}_s$, so that, from (\ref{eq9.2}),
${\rm min}[\delta_0(\hat{x})]=\delta_0(\hat{x}_s)=-(3\sqrt{3}/8)(u_m/a)$. Application
of (\ref{eq9.6}) along the characteristic $\hat{x}=\hat{x}_s$ yields
\begin{equation}                                    
        \left(\frac{\delta^2(x,t) + \zeta^2(x,t)}{\delta_0^2(\hat{x}_s)}\right)^{1/2}
     = \frac{e^{-kt}}{1 - (3\sqrt{3}/8)(u_m/a)t_1(t)}.
\label{eq9.7}
\end{equation}
Thus, the shock formation time $t_s$ is given implicitly by
\begin{equation}                                  
         t_1(t_s) = \frac{8\sqrt{3}}{9}\frac{a}{u_m} \approx 1.54\,\frac{a}{u_m},
\label{eq9.8}
\end{equation}
and, from (\ref{eq9.4}), the position of shock formation is
\begin{equation}                                  
             x_s = -\sqrt{3}\, a + u_{_E} t_s.
\label{eq9.9}
\end{equation}
Note from Fig.~16 that equation (\ref{eq9.8}) has a solution only when $(8\sqrt{3}/9)(a/u_m)$
is smaller than the maximum value of $t_1(t)$. The maximum value of $t_1(t)$
occurs at $t=\pi/(2f) \approx 8.73$ h and, from (\ref{eq7.11}), has the value
   $$  {\rm max}[t_1(t)] = \frac{k+f\exp[-(\pi/2)(k/f)]}{f^2+k^2}.  $$
Thus, the condition for shock formation is
\begin{equation}                                  
         \frac{u_m}{fa} > \left(\frac{u_m}{fa}\right)_c
       \equiv \frac{8\sqrt{3}}{9}\left(\frac{1 + (k/f)^2}{(k/f) + \exp[-(\pi/2)(k/f)]}\right).
\label{eq9.10}
\end{equation}

\begin{figure*}[!t]                    
\centerline{\includegraphics[width=39pc]{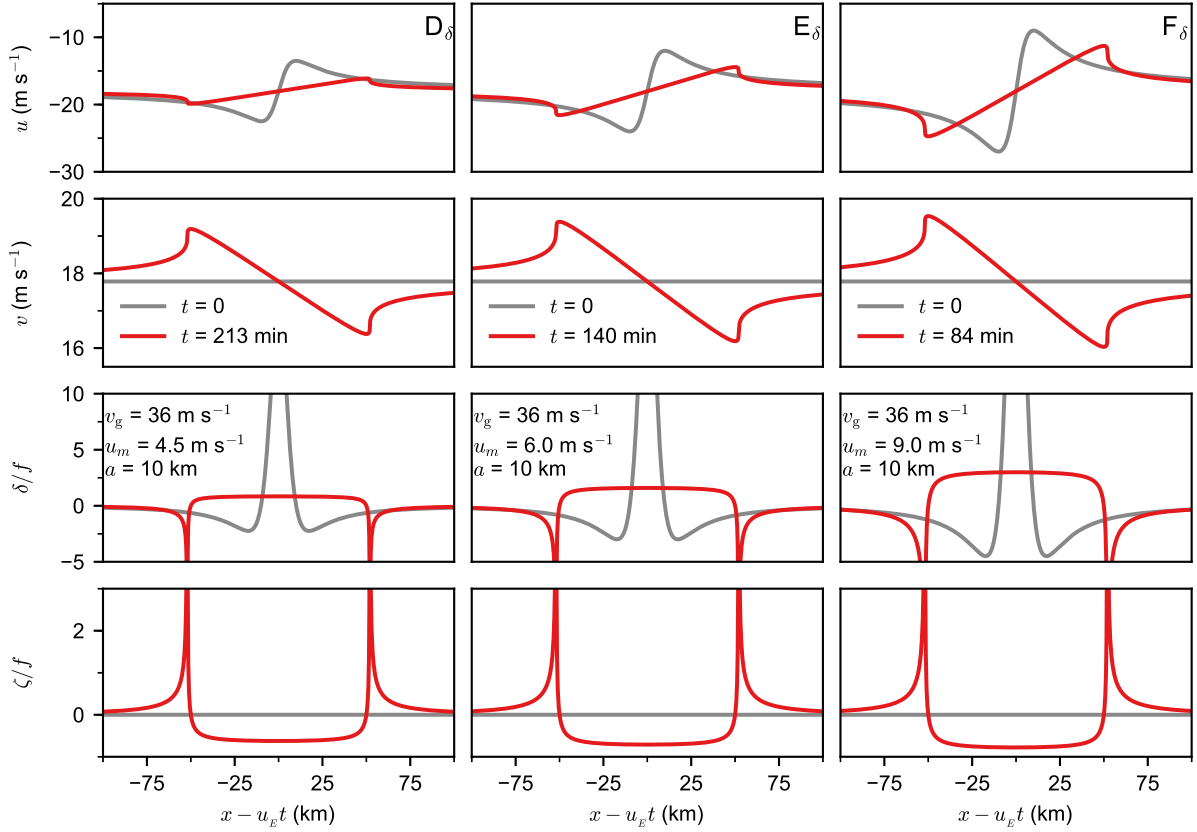}}
\caption{The three columns show three examples with initial divergence only,
as determined by the analytical solutions (\ref{eq9.13})--(\ref{eq9.15}). All three
lead to the formation of N-waves in $u$ and $v$. The
spatial distributions at $t=0$ are shown by the gray curves, while the distributions
at shock formation time are shown by the red curves. All three cases have
$v_g=36$ m s$^{-1}$, $a=10$ km, and the same initial $v_0(x)=v_{_E}=17.8$ m~s$^{-1}$,
so that the initial vorticity is zero. These three examples correspond to the three
points labeled $D_\delta,E_\delta,F_\delta$ in the bottom panel of Fig.~17.}
\end{figure*}

     Equations (\ref{eq9.8}) and (\ref{eq9.10}) have been used to construct
the top panel of Fig.~17, which shows isolines of the shock formation time
$t_s$ and the shock critical condition (thick line) in the $(u_m/fa,v_g)$-plane.
There is only a weak dependence of $t_s$ on $v_g$, with shock formation times
less than one hour when $u_m/fa > 9$. The three columns of Fig.~18 show three
examples of the analytical solutions (\ref{eq9.3})--(\ref{eq9.5}). The
four rows show plots of $u$, $v$, $\delta/f$, and $\zeta/f$ as functions of
$x-u_{_E} t$ with the initial conditions given by the gray curves and
the distributions at shock formation time given by the red curves (corresponding
to 45 min for the left column, 29 min for the middle column, and 22 min for
the right column). As the spatial variation of the initial $u$ increases, the
final jump in $u$ also increases, but the final jump in $v$ changes little.
Since the final jumps in $v$ are smaller than the corresponding final jumps
in $u$, all three cases can be classified as divergence-preferred triangular
waves.

\subsection{Formation of an N-wave}

     For the second example, consider the initial conditions
\begin{equation}                                  
       u_0(x) = u_{_E} + u_m \left(\frac{2x/a}{1 + (x/a)^2}\right)
       \quad \text{and} \quad    v_0(x) = v_{_E},
\label{eq9.11}
\end{equation}
where the constants $a$ and $u_m$ now specify the horizontal extent and
strength of this initial antisymmetric divergent flow anomaly.  The initial
divergence and vorticity associated with (\ref{eq9.11}) are
\begin{equation}                                  
     \delta_0(x) = \frac{2u_m}{a} \left(\frac{1 - (x/a)^2}{[1 + (x/a)^2]^2}\right),
     \quad \text{and} \quad   \zeta_0(x)  = 0.
\label{eq9.12}
\end{equation}
We assume $u_m>0$, which is the case leading to an N-wave and a double shock. With
the initial conditions (\ref{eq9.11}), the solutions (\ref{eq7.8}) and (\ref{eq7.9})
become
\begin{equation}                                  
  \begin{split}
       u(x,t) &= u_{_E} + u_m\left(\frac{2\hat{x}/a}{1 + (\hat{x}/a)^2}\right)e^{-kt}\cos(ft), \\
       v(x,t) &= v_{_E} - u_m\left(\frac{2\hat{x}/a}{1 + (\hat{x}/a)^2}\right)e^{-kt}\sin(ft),
  \end{split}
\label{eq9.13}
\end{equation}
and the characteristic equation (\ref{eq7.10}) becomes
\begin{equation}                                  
       x = \hat{x} + u_{_E}t + u_m\left(\frac{2\hat{x}/a}{1 + (\hat{x}/a)^2}\right)t_1(t).
\label{eq9.14}
\end{equation}
The solutions (\ref{eq7.13}) for the divergence and vorticity become
\begin{equation}                                  
      \delta(x,t) = \frac{\delta_0(\hat{x})e^{-kt}\cos(ft)}{1 + t_1(t)\delta_0(\hat{x})}, \quad
      \zeta (x,t) =-\frac{\delta_0(\hat{x})e^{-kt}\sin(ft)}{1 + t_1(t)\delta_0(\hat{x})},
\label{eq9.15}
\end{equation}
so that
\begin{equation}                                    
        \left(\frac{\delta^2(x,t) + \zeta^2(x,t)}{\delta_0^2(\hat{x})}\right)^{1/2}
     = \frac{e^{-kt}}{1 + t_1(t)\delta_0(\hat{x})}.
\label{eq9.16}
\end{equation}
From (\ref{eq9.15}) or (\ref{eq9.16}), shock formation occurs along a characteristic
$\hat{x}$ when $t_1(t)\delta_0(\hat{x}) = -1$. This occurs first along the two characteristics
with the minimum value of the initial divergence $\delta_0(\hat{x})$, i.e., along the
two characteristics with the maximum value of the initial convergence. For this example,
the two minimum values of $\delta_0(\hat{x})$ occur at $\hat{x}=\pm\sqrt{3}\, a$, so that,
from (\ref{eq9.12}), ${\rm min}[\delta_0(\hat{x})] = -u_m/4a$. Application of (\ref{eq9.16})
along the two characteristics $\hat{x}=\pm\hat{x}_s$, where $\hat{x}_s=\sqrt{3}\, a$, yields
\begin{equation}                                    
        \left(\frac{\delta^2(x,t) + \zeta^2(x,t)}{\delta_0^2(\pm\hat{x}_s)}\right)^{1/2}
     = \frac{e^{-kt}}{1 - (u_m/4a)t_1(t)}.
\label{eq9.17}
\end{equation}
Thus, the shock formation time $t_s$ is given implicitly by
\begin{equation}                                  
          t_1(t_s) = \frac{4a}{u_m},
\label{eq9.18}
\end{equation}
and, from (\ref{eq9.14}), the positions of shock formation are
\begin{equation}                                  
            x_s = u_{_E} t_s \pm 3\sqrt{3}\, a.
\label{eq9.19}
\end{equation}
From Fig.~16, equation (\ref{eq9.18}) has a solution only when $4a/u_m$
is smaller than the maximum value of $t_1(t)$. The maximum value of $t_1(t)$
occurs at $t=\pi/(2f)\approx 8.73$ h and, from (\ref{eq7.11}), has the value
   $$  {\rm max}[t_1(t)] = \frac{k+f\exp[-(\pi/2)(k/f)]}{f^2+k^2}.   $$
Thus, the condition for shock formation is
\begin{equation}                                  
       \frac{u_m}{4fa} > \left(\frac{u_m}{4fa}\right)_c
       \equiv \frac{1 + (k/f)^2}{(k/f) + \exp[-(\pi/2)(k/f)]}.
\label{eq9.20}
\end{equation}

     Isolines of the shock formation time $t_s$, as given implicitly by
(\ref{eq9.18}), and the shock condition, as given by (\ref{eq9.20}), are
shown in the bottom panel of Fig.~17.  The solutions for $u,v,\delta/f,\zeta/f$,
as given by (\ref{eq9.13})--(\ref{eq9.15}), are plotted in Fig.~19
for the particular constants $v_g=36$ m s$^{-1}$, $a=10$ km, and for
the three cases $u_m=4.5,\, 6.0,\, 9.0$ m s$^{-1}$. All three examples
evolve into N-waves in $u$ and $v$, and therefore singularities in the
Ekman pumping on both edges of the widening moat.

The time evolution of the divergence and vorticity along the shock-producing
characteristics for these N-waves is shown by the bluish curves in the
lower panel of Fig.~20. All three cases are divergence-preferred,
so the discontinuities in $u$ are larger than those in $v$.

\begin{figure}[!thp]               
\centerline{\includegraphics[width=19pc]{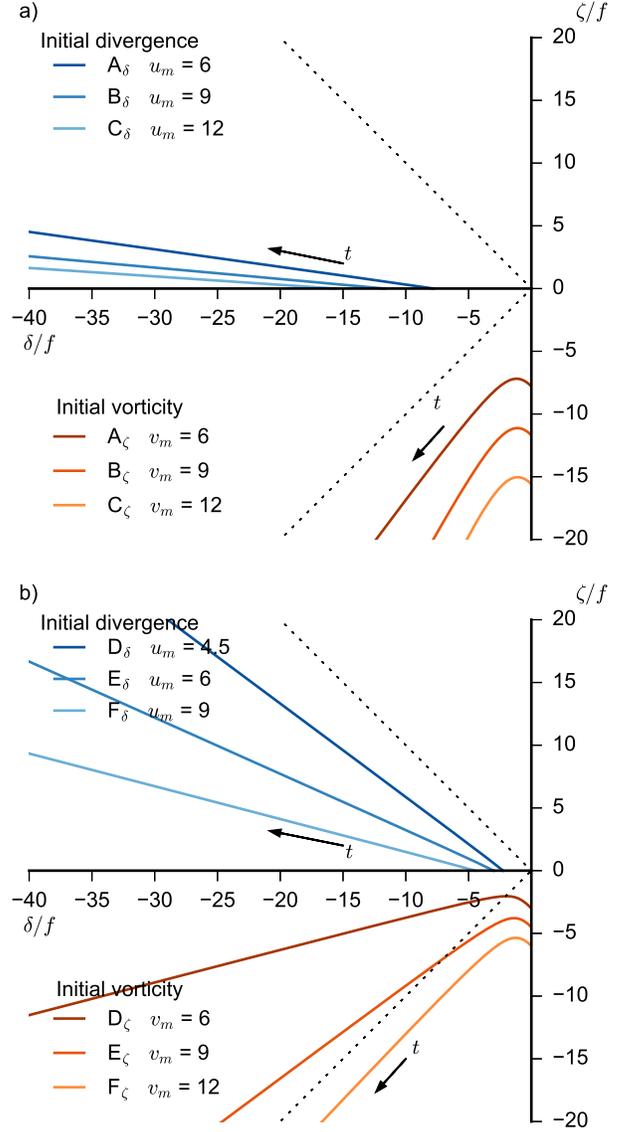}}
\caption{Time evolution of the vorticity and divergence along the first shock-producing
characteristics for the triangular waves described in sections 9a and 10a (upper panel)
and for the N-waves described in sections 9b and 10b (lower panel). The bluish curves
are initialized with $\zeta=0$ and $\delta\ne0$, while the reddish curves are
initialized with $\delta=0$ and $\zeta\ne0$. The direction of increasing time
is indicated by the arrows. The dashed lines are defined by $|\zeta|=|\delta|$.
In the upper panel, the three cases A$_\delta$, B$_\delta$, C$_\delta$ are
divergence preferred triangular waves, while the three cases A$_\zeta$, B$_\zeta$,
C$_\zeta$ are vorticity preferred triangular waves. In the lower panel, the
four cases D$_\delta$, E$_\delta$, F$_\delta$, D$_\zeta$ are divergence-preferred
N-waves, while E$_\zeta$ and F$_\zeta$ are N-waves with nearly the same
magnitude in the singularities of $\delta$ and $\zeta$.}
\end{figure}

\begin{figure}[!th]                          
\centerline{\includegraphics[width=19pc]{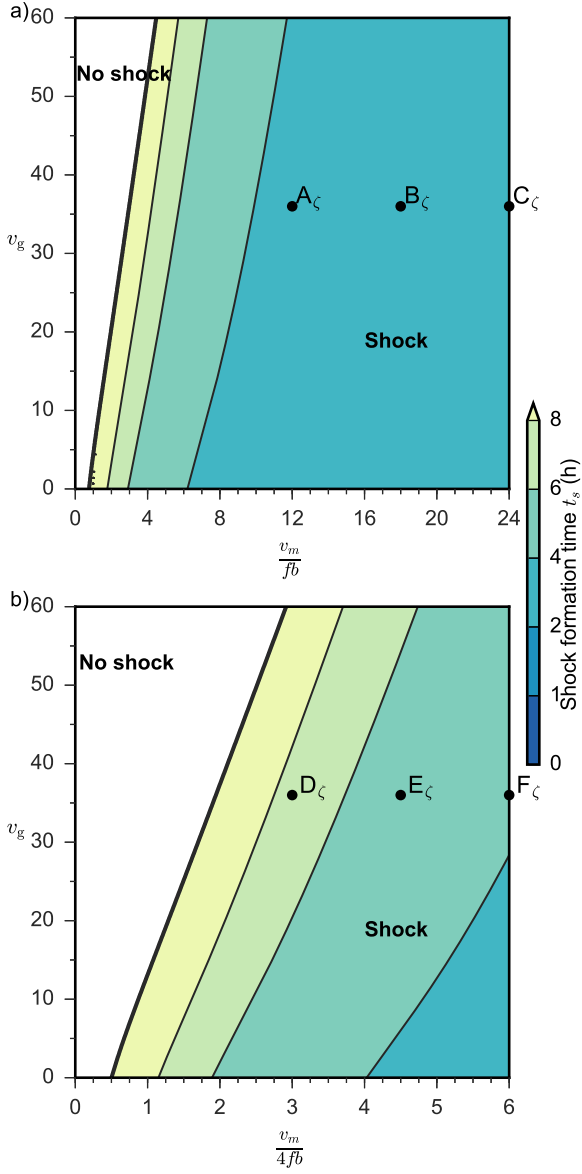}}
\caption{Isolines of the shock formation time $t_s$ (color shading) and the shock
condition (thick curve) for a) the initial vorticity cases resulting in triangular
waves, as described by equations (\ref{eq10.8}) and (\ref{eq10.10}) and b) the
initial vorticity cases resulting in N-waves, as described by
equations (\ref{eq10.28}) and (\ref{eq10.30}). The abscissa in each panel is
a dimensionless measure of the maximum initial vorticity. The points
$A_\zeta,B_\zeta,C_\zeta$ in the upper panel correspond to the three columns
in Fig.~22, while the points $D_\zeta,E_\zeta,F_\zeta$ in the lower panel
correspond to the three columns in Fig.~23.}
\end{figure}

\section{Examples with initial vorticity only}     

     In this section, it is shown that triangular waves and N-waves can also be
produced from initial conditions that have zero divergence and nonzero vorticity.

\subsection{Formation of a triangular wave}

     As the first set of simple examples for this section, consider the initial
conditions
\begin{equation}                                  
       u_0(x) = u_{_E}  \quad \text{and} \quad v_0(x) = v_{_E} - v_m \left(\frac{1}{1 + (x/b)^2}\right),
\label{eq10.1}
\end{equation}
where the constants $b$ and $v_m$ specify the horizontal extent and strength of this
initial rotational flow anomaly. The initial divergence and vorticity associated
with (\ref{eq10.1}) are
\begin{equation}                                  
     \delta_0(x) = 0  \quad \text{and} \quad
     \zeta_0(x)  = \frac{2v_m}{b} \left(\frac{x/b}{[1 + (x/b)^2]^2}\right).
\label{eq10.2}
\end{equation}
We assume $v_m>0$ so that negative initial vorticity appears to the left
of the origin. With these initial conditions, the solutions (\ref{eq7.8})
and (\ref{eq7.9}) simplify to
\begin{equation}                                  
  \begin{split}
       u(x,t) &= u_{_E} - v_m\left(\frac{1}{1 + (\hat{x}/b)^2}\right)e^{-kt}\sin(ft), \\
       v(x,t) &= v_{_E} - v_m\left(\frac{1}{1 + (\hat{x}/b)^2}\right)e^{-kt}\cos(ft),
  \end{split}
\label{eq10.3}
\end{equation}
while the characteristic equation (\ref{eq7.10}) simplifies to
\begin{equation}                                  
       x = \hat{x} + u_{_E}t - v_m\left(\frac{1}{1 + (\hat{x}/b)^2}\right)t_2(t).
\label{eq10.4}
\end{equation}
The solutions (\ref{eq7.13}) for the divergence and vorticity become
\begin{equation}                                  
       \delta(x,t) = \frac{\zeta_0(\hat{x})e^{-kt}\sin(ft)}{1 + t_2(t)\zeta_0(\hat{x})},   \quad
       \zeta (x,t) = \frac{\zeta_0(\hat{x})e^{-kt}\cos(ft)}{1 + t_2(t)\zeta_0(\hat{x})},
\label{eq10.5}
\end{equation}
so that
\begin{equation}                                    
        \left(\frac{\delta^2(x,t) + \zeta^2(x,t)}{\zeta_0^2(\hat{x})}\right)^{1/2}
     = \frac{e^{-kt}}{1 + t_2(t)\zeta_0(\hat{x})}.
\label{eq10.6}
\end{equation}
From (\ref{eq10.5}) or (\ref{eq10.6}), shock formation occurs along the characteristic
$\hat{x}$ when $t_2(t)\zeta_0(\hat{x}) = -1$. This occurs first along the characteristic
with the minimum value of $\zeta_0(\hat{x})$. For this example, the minimum value of
$\zeta_0(\hat{x})$ occurs at $\hat{x}=-b/\sqrt{3}\equiv\hat{x}_s$ so that, from (\ref{eq10.2}),
${\rm min}[\zeta_0(\hat{x})]=\zeta_0(\hat{x}_s) = -(3\sqrt{3}/8)(v_m/b)$. Application
of (\ref{eq10.6}) along the characteristic $\hat{x}=\hat{x}_s$ yields
\begin{equation}                                    
        \left(\frac{\delta^2(x,t) + \zeta^2(x,t)}{\zeta_0^2(\hat{x}_s)}\right)^{1/2}
     = \frac{e^{-kt}}{1 - (3\sqrt{3}/8)(v_m/b)t_2(t)}.
\label{eq10.7}
\end{equation}
Thus, the shock formation time $t_s$ is given implicitly by
\begin{equation}                                  
         t_2(t_s) = \frac{8\sqrt{3}}{9}\frac{b}{v_m} \approx 1.54\,\frac{b}{v_m},
\label{eq10.8}
\end{equation}
and, from (\ref{eq10.4}), the position of shock formation is
\begin{equation}                                  
             x_s = u_{_E} t_s - \sqrt{3}\, b.
\label{eq10.9}
\end{equation}
Note from Fig.~16 that equation (\ref{eq10.8}) has a solution only when $(8\sqrt{3}/9)(b/v_m)$
is smaller than the maximum value of $t_2(t)$. The maximum value of $t_2(t)$
occurs at $t=\pi/f \approx 17.46$ h and, from (\ref{eq7.11}), has the value
   $$  {\rm max}[t_2(t)] = \frac{f+f\exp[-\pi(k/f)]}{f^2+k^2}.  $$
Thus, the condition for shock formation is
\begin{equation}                                  
         \frac{v_m}{fb} > \left(\frac{v_m}{fb}\right)_c
       \equiv \frac{8\sqrt{3}}{9}\left(\frac{1 + (k/f)^2}{1 + \exp[-\pi(k/f)]}\right).
\label{eq10.10}
\end{equation}
\begin{figure*}[t]                   
\centerline{\includegraphics[width=39pc]{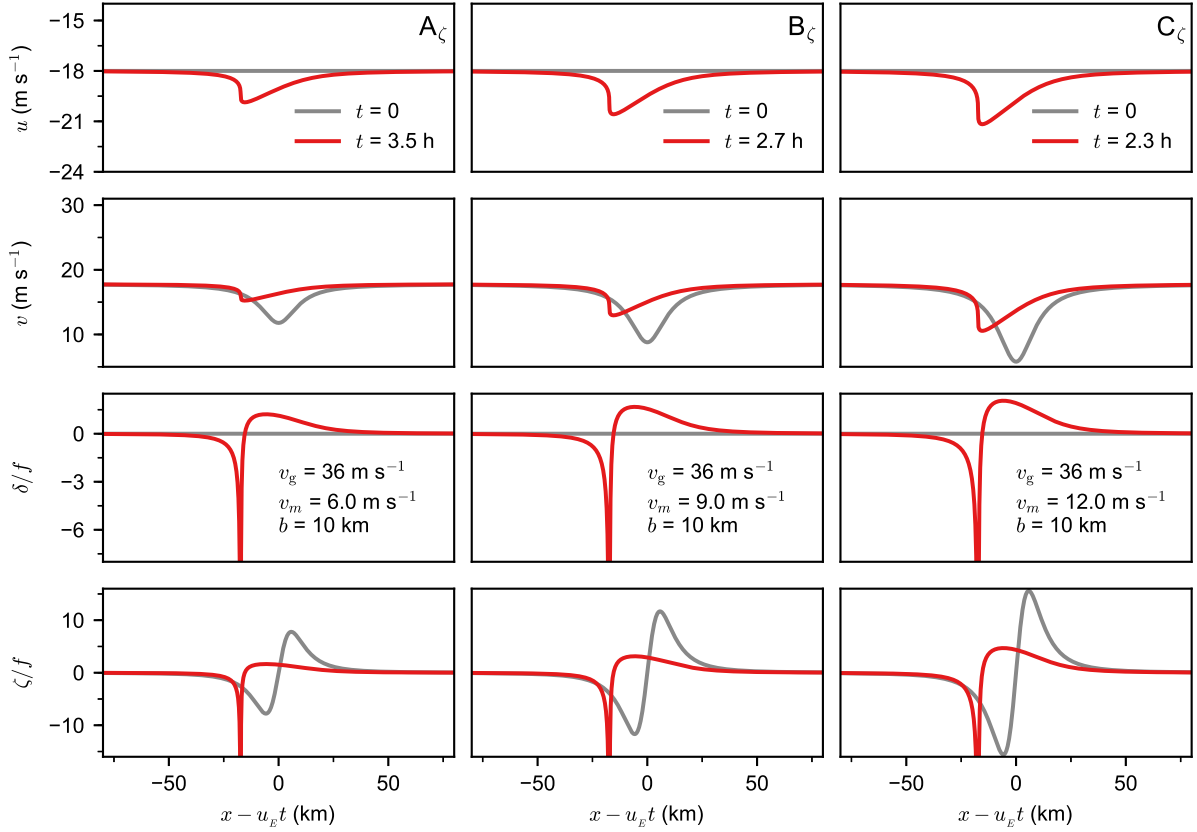}}
\caption{The three columns show three examples with initial vorticity only,
as determined by the analytical solutions (\ref{eq10.3})--(\ref{eq10.5}). All three
lead to the formation of triangular waves. The spatial distributions
at $t=0$ are shown by the gray curves, while the distributions
at shock formation time are shown by the red curves. All three cases have
$v_g=36$ m s$^{-1}$, $b=10$ km, and the same initial $u_0(x)=u_{_E}=-18$ m~s$^{-1}$,
so that the initial divergence is zero.}
\end{figure*}

The solutions for
$u$, $v$, $\delta/f$, and $\zeta/f$, as given by (\ref{eq10.3})--(\ref{eq10.5}),
are plotted in Fig.~22 for the constants $v_g=36$ m s$^{-1}$,
$b=10$ km, and $v_m=6,9,12$ m s$^{-1}$ for cases A$_\zeta$, B$_\zeta$, and C$_\zeta$.
The plots cover the spatial interval $-80 \le x \le 80$ km and are for $t=0$ and
$t=t_s$, where $t_s=3.5, 2.7, 2.3$ h is the shock formation
time for each initial condition.

     The time evolution of the divergence and vorticity along the first
shock-producing characteristic is shown by the reddish curves in the top
panel of Fig.~20. Note from (\ref{eq10.5}) that
$\delta/\zeta = \tan(ft)$, so that the shock is vorticity-preferred if $0 < ft_s < \pi/4$
and is divergence-preferred if $\pi/4 < ft_s < \pi/2$.
All three of the cases A$_\zeta$, B$_\zeta$, C$_\zeta$ fall in the former
range and are therefore vorticity-preferred triangular waves.

\bigskip

\subsection{Formation of an N-wave}
\begin{figure*}[t]                    
\centerline{\includegraphics[width=39pc]{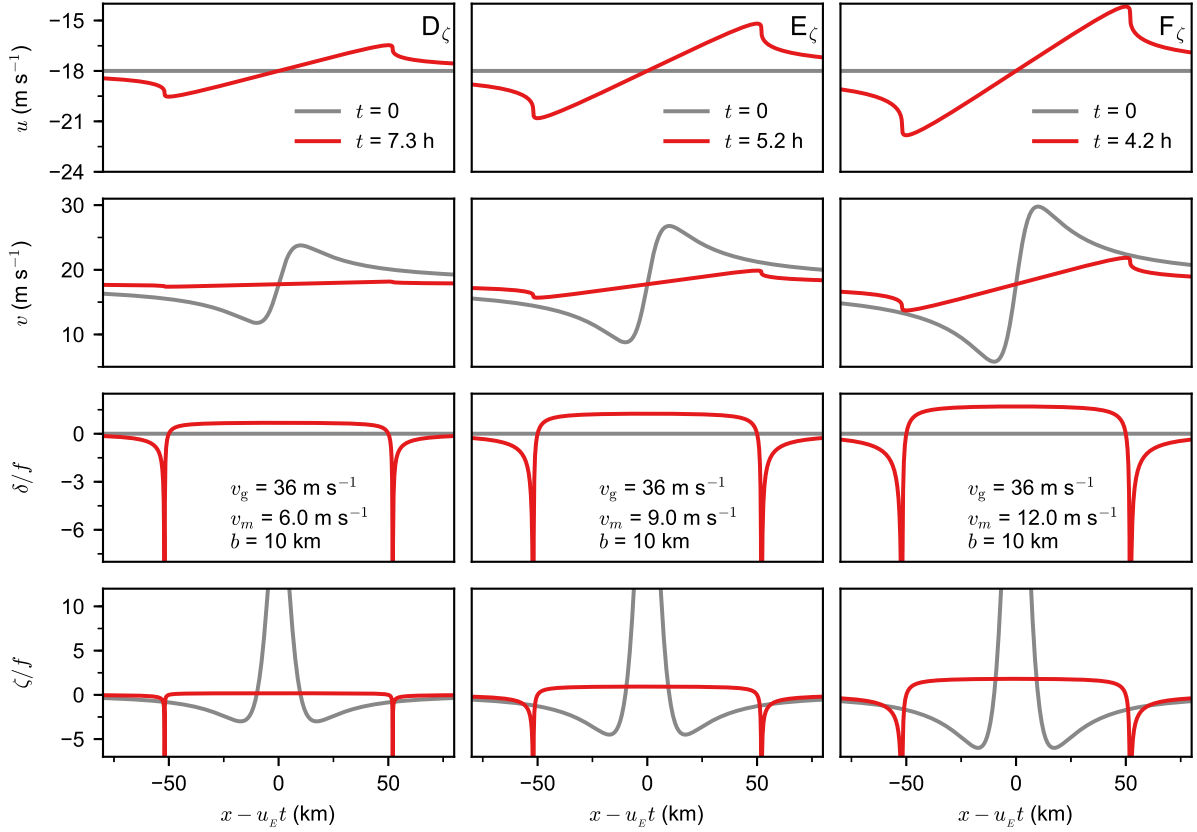}}
\caption{The three columns show three examples with initial vorticity only,
as determined by the analytical solutions (\ref{eq10.23})--(\ref{eq10.25}). The
spatial distributions at $t=0$ are shown by the gray curves, while the distributions
at shock formation time are shown by the red curves. All three cases have
$v_g=36$ m s$^{-1}$, $a=10$ km, and the same initial $u_0(x)=u_{_E}=18$ m~s$^{-1}$,
so that the initial divergence is zero.}
\end{figure*}

     As the second set of simple examples for this section, consider the initial
conditions
\begin{equation}                                  
       u_0(x) = u_{_E}  \quad \text{and} \quad v_0(x) = v_{_E} + v_m \left(\frac{2x/b}{1 + (x/b)^2}\right),
\label{eq10.21}
\end{equation}
where the constants $b$ and $v_m$ now specify the horizontal extent and strength
of this initial anti-symmetric rotational flow anomaly. The initial divergence
and vorticity associated with (\ref{eq10.21}) are
\begin{equation}                                  
     \delta_0(x) = 0  \quad \text{and} \quad
     \zeta_0(x)  = \frac{2v_m}{b} \left(\frac{1-(x/b)^2}{[1+(x/b)^2]^2}\right).
\label{eq10.22}
\end{equation}
We assume $v_m>0$, so that negative initial vorticity appears on the wings
of a central region of positive vorticity. With these initial conditions, the
solutions (\ref{eq7.8}) and (\ref{eq7.9}) simplify to
\begin{equation}                                  
  \begin{split}
       u(x,t) &= u_{_E} + v_m\left(\frac{2\hat{x}/b}{1 + (\hat{x}/b)^2}\right)e^{-kt}\sin(ft), \\
       v(x,t) &= v_{_E} + v_m\left(\frac{2\hat{x}/b}{1 + (\hat{x}/b)^2}\right)e^{-kt}\cos(ft),
  \end{split}
\label{eq10.23}
\end{equation}
while the characteristic equation (\ref{eq7.10}) simplifies to
\begin{equation}                                  
       x = \hat{x} + u_{_E}t + v_m\left(\frac{2\hat{x}/b}{1 + (\hat{x}/b)^2}\right)t_2(t).
\label{eq10.24}
\end{equation}
The solutions (\ref{eq7.13}) for the divergence and vorticity become
\begin{equation}                                  
       \delta(x,t) = \frac{\zeta_0(\hat{x})e^{-kt}\sin(ft)}{1 + t_2(t)\zeta_0(\hat{x})},   \quad
       \zeta (x,t) = \frac{\zeta_0(\hat{x})e^{-kt}\cos(ft)}{1 + t_2(t)\zeta_0(\hat{x})},
\label{eq10.25}
\end{equation}
so that
\begin{equation}                                    
        \left(\frac{\delta^2(x,t) + \zeta^2(x,t)}{\zeta_0^2(\hat{x})}\right)^{1/2}
     = \frac{e^{-kt}}{1 + t_2(t)\zeta_0(\hat{x})}.
\label{eq10.26}
\end{equation}
From (\ref{eq10.25}) or (\ref{eq10.26}), shock formation occurs along a characteristic
$\hat{x}$ when $t_2(t)\zeta_0(\hat{x}) = -1$. This occurs first along the two characteristics
with the minimum value of the initial vorticity $\zeta_0(\hat{x})$. For this example,
the two minimum values of $\zeta_0(\hat{x})$ occur at $\hat{x}=\pm\sqrt{3}\, b$, so that,
from (\ref{eq10.22}), ${\rm min}[\zeta_0(\hat{x})] = -v_m/4b$. Application of (\ref{eq10.26})
along the two characteristics $\hat{x}=\pm\hat{x}_s$, where $\hat{x}_s=\sqrt{3}\, b$, yields
\begin{equation}                                    
        \left(\frac{\delta^2(x,t) + \zeta^2(x,t)}{\zeta_0^2(\pm\hat{x}_s)}\right)^{1/2}
     = \frac{e^{-kt}}{1 - (v_m/4b)t_2(t)}.
\label{eq10.27}
\end{equation}
Thus, the shock formation time $t_s$ is given implicitly by
\begin{equation}                                  
          t_2(t_s) = \frac{4b}{v_m},
\label{eq10.28}
\end{equation}
and, from (\ref{eq10.24}), the positions of shock formation are
\begin{equation}                                  
            x_s = u_{_E} t_s \pm 3\sqrt{3}\, b.
\label{eq10.29}
\end{equation}
From Fig.~16, equation (\ref{eq10.28}) has a solution only when $4b/v_m$
is smaller than the maximum value of $t_2(t)$. The maximum value of $t_2(t)$
occurs at $t=\pi/f\approx 17.46$ h and, from (\ref{eq7.11}), has the value
   $$   {\rm max}[t_2(t)] = \frac{f+f\exp[-\pi(k/f)]}{f^2+k^2}.  $$
Thus, the condition for shock formation is
\begin{equation}                                  
       \frac{v_m}{4fb} > \left(\frac{v_m}{4fb}\right)_c
       \equiv \frac{1 + (k/f)^2}{1 + \exp[-\pi(k/f)]}.
\label{eq10.30}
\end{equation}

The solutions for $u$, $v$, $\delta/f$, $\zeta/f$, as given by \eqref{eq10.23}--\eqref{eq10.25},
are plotted in Fig.~23, using the constants $v_g=36$ m s$^{-1}$,
$b=$ 10 km, and $v_m=6,9,12$ m s$^{-1}$ for cases D$_\zeta$, E$_\zeta$, F$_\zeta$.
As can be seen from the lower panel of Fig.~20, the case D$_\zeta$ produces
divergence-preferred N-wave shocks, while cases E$_\zeta$ and F$_\zeta$
produce N-wave shocks that are of nearly equal strength in divergence and vorticity.

\section{Concluding remarks}          

    In sections 2--6, we have reviewed the theory of the one-dimensional
nonlinear advection equation and Burgers' equation.  These two equations
provide a simple framework for understanding the concepts of triangular
waves and N-waves. In sections 7--10, we have considered the line-symmetric
slab boundary layer model (\ref{eq7.1}). Although this model lacks important
curvature effects that are present in the axisymmetric slab boundary layer
model (\ref{eq1.1}), the line-symmetric model is simple enough for analytical
progress and generalization of the concepts of triangular waves and N-waves.
In particular, the line-symmetric model (\ref{eq7.1}) has been solved by
taking advantage of its
hyperbolic form, thereby rewriting it as the system of three ordinary
differential equations given in (\ref{eq7.7}). The solutions of these
three ordinary differential equations are given in (\ref{eq7.8})--(\ref{eq7.11}),
with the associated divergence and vorticity solutions given in (\ref{eq7.13}).
When the denominators on the right hand sides of (\ref{eq7.13}) vanish,
the divergence and vorticity become infinite, so there appears a singularity
in the boundary layer pumping along with a vertically oriented vorticity sheet
in the boundary layer. As shown by the examples in sections 9 and 10, such shocks
develop when the initial convergence or initial vorticity exceed a critical
value. The shocks can be classified as divergence-preferred or
vorticity-preferred, depending on whether the jump in the divergent component
$u$ is larger than the jump in the rotational component $v$, or vice versa.
In this regard it is interesting to note that the classic Hurricane Hugo case
(Fig.~1) can be interpreted as a vorticity-preferred
shock, with the jump in the rotational component nearly three times as large
as the jump in the divergent component.

    The plots shown in sections 9 and 10 display the solutions up to the
shock formation time $t=t_s$. How do we extend the solutions beyond $t_s$,
i.e., into regions of the $(x,t)$-plane where characteristics intersect and
the solutions (\ref{eq7.8}) and (\ref{eq7.9}) become multivalued? Although
the momentum
equations (\ref{eq7.1}) remain valid in the smooth regions of flow, these
equations are not useful at the discontinuity, where $(\partial u/\partial x)$
and $(\partial v/\partial x)$ become infinite. Thus, equations (\ref{eq7.1}) need
to be supplemented by jump conditions that describe the dynamics across the shock.
One practical alternative to the use of jump conditions is to include horizontal
diffusion terms in (\ref{eq7.1}), but to set the diffusivity constant to such
a small value that the horizontal diffusion terms have importance only in the
region near the shock. Then, since $(\partial u/\partial x)$ and
$(\partial v/\partial x)$ are prevented from becoming infinite, explicit jump
conditions are not required. This strategy of including horizontal diffusion was
used in the numerical simulations of single and double eyewalls shown in Fig.~4.
Another practical alternative to the use of jump
conditions involves shock-capturing numerical methods such as those used by
\citet{kuo+polvani97} to simulate the shocks that appear as transient features
in the fully nonlinear, shallow-water, geostrophic adjustment problem.

     The results presented here provide some insight into questions such
as: (1) What determines the size of the eye? (2) How are potential vorticity
rings produced?  (3) How does an outer concentric eyewall form and how does it
influence the inner eyewall?
The slab boundary layer results support the notion that the size of the eye is
determined by nonlinear processes that set the radius at which the eyewall
shock forms.  A boundary layer potential vorticity ring is also produced at
this radius. By boundary layer pumping and latent heat release, the boundary
layer potential vorticity ring is extended upward.  If, outside the eyewall, the boundary layer
radial inflow does not decrease monotonically with radius, a concentric eyewall
boundary layer shock can form. If it is strong enough and close enough to the
inner eyewall, this outer eyewall shock can chock off the boundary layer radial
inflow to the inner shock and effectively shut down the boundary layer pumping
at the inner eyewall. An important issue not explored here is the distinction
between weak and strong shocks. Although some initial conditions can technically
produce shocks, they may be too weak to be of physical significance.
Thus, an interesting remaining problem is to determine the conditions
that produce strong enough shocks that the boundary layer can take
control of the organization of the deep moist convection in the cyclone.

    The frictional boundary layer comprises only about 10\% of the mass involved
in the tropical cyclone circulation. However, because of its high moisture content
and its tendency to produce regions of intense convergence and Ekman pumping, it
can dictate the location and strength of primary and secondary eyewalls. Thus,
the dynamical importance of the frictional boundary layer far exceeds its
fractional mass content. Concerning the location and strength of eyewall
convection, the basic idea presented here is that the single and double eyewall
structures observed in tropical cyclones are a result of the nonlinear
dynamics of the boundary layer. Because of the $u(\partial u/\partial r)$
term in the radial equation of motion, divergent regions of boundary layer
flow broaden and weaken with time, while convergent regions sharpen and
strengthen with time. Single eyewalls develop when the sharpening process
is dominant on the inside edge of the divergent moat, in analogy with the
development of a triangular wave. Concentric eyewalls develop when the
sharpening process is active on both sides of the moat, in analogy with
the development of an N-wave. An interesting aspect of tropical cyclone
boundary layer dynamics is that the $u(\partial u/\partial r)$ term in
the radial equation of motion can be negligible at most radii, so that
a local Ekman theory (i.e., a theory that neglects radial advection)
yields a reasonable approximation to the flow at most radii. However,
as a tropical cyclone intensifies, there can develop radial intervals
where local Ekman theory breaks down and Burgers-type sharpening effects
become dominant in determining the boundary layer flow structure. It can
be argued that such sharpening processes are crucial in producing the
classic single or double eyewall structures that define a hurricane.

    In closing we note that the present study has focused on understanding
the boundary layer response to a specified, axisymmetric, non-translating pressure
field, in which case the boundary layer shocks are circular. Tropical cyclones
are rarely stationary, and when the pressure field translates, the boundary
layer shocks can form more complicated structures, such as crescent shapes or
spiral shapes. Understanding the development of boundary layer shocks
forced by a translating pressure field remains a challenging problem.

\begin{acknowledgment}
     We would like to thank Alex Gonzalez, Paul Ciesielski, and Gabriel Williams
for their advice. This work has been supported by the NSF under Grants AGS-1546610
and AGS-1601623 and by the Hurricane Forecast Improvement Project (HFIP) under
NOAA Grant NA090AR4320074.
\end{acknowledgment}

\ifthenelse{\boolean{dc}}
{}
{\clearpage}
\bibliographystyle{ametsoc2014}
\bibliography{references}

\end{document}